\documentclass[printer]{aa}
\usepackage{lineno}
\usepackage[normalem]{ulem}
\usepackage{textcomp}
\usepackage{float}
\usepackage{dcolumn}
\newcolumntype{d}[1]{D{.}{.}{#1}}  
\usepackage{graphicx}
\usepackage{xcolor,colortbl}
\usepackage{booktabs}
\usepackage{txfonts}
\usepackage{subfigure}
\usepackage{xcolor}
\usepackage{amsmath}
\usepackage[colorlinks=true,urlcolor=blue,citecolor=blue,linkcolor=blue,breaklinks=true]{hyperref}
\begin{document}

\title{A molecule-rich torus-like structure in the  
21\,$\mu$m source IRAS~23304+6147}

\titlerunning{Torus-like structure  in IRAS~23304+6147}
\authorrunning{Sun et al.}


\author{Hao-Min Sun\inst{1}
          \and
          Yong Zhang
          \inst{1,2,3\thanks{zhangyong5@email.sysu.edu.cn}}
          \and
          Xu-Jia Ouyang\inst{1}
          \and
          Sheng-Li Qin\inst{4}
          \and
          Junichi Nakashima\inst{1,3}
          \and
          Jian-Jie Qiu\inst{5}
          \and
          Xiao-Hu Li\inst{2,6,7}
          }

\institute{School of Physics and Astronomy, Sun Yat-sen University, 2 Daxue Road, Tangjia, Zhuhai 519082, Guangdong Province, China \email{zhangyong5@email.sysu.edu.cn}
      \and
          Xinjiang Astronomical Observatory, Chinese Academy of Sciences, 150 Science 1-Street, Urumqi 830011, China
      \and
          CSST Science Center for the Guangdong-Hongkong-Macau Greater Bay Area, Sun Yat-Sen University, Guangdong Province, China
      \and
          School of physics and astronomy, Yunnan University, Kunming 650091, China
      \and
          School of Mathematics and Physics, Jinggangshan University, 28 Xueyuan Road, Qingyuan District, Ji’an 343009, Jiangxi Province, China
        \and 
        Xinjiang Key Laboratory of Radio Astrophysics, 150 Science 1-Street, Urumqi, Xinjiang 830011, China
        \and
        Key Laboratory of Radio Astronomy and Technology, Chinese Academy of Sciences, A20 Datun Road, Chaoyang District, Beijing, 100101, China
          }

   \date{\today}

 
\abstract
{A long-standing enigma in observational astronomy is the identification of the so-called 21\,$\mu$m feature in a subset of envelopes of post-asymptotic giant branch (post-AGB) stars.
Identifying this transient feature is important for understanding the chemical processes during the brief post-AGB phase and the enrichment of the interstellar medium. 
Understanding the structures and chemical environments of these objects is a prerequisite for such an endeavor. }
{ We investigate the structure of the circumstellar envelope and the spatial distribution of gas-phase molecules in the 21\,$\mu$m source IRAS~23304+6147, aiming to explor the potential physicochemical conditions required for the emergence of the 21\,$\mu$m feature.}
{Molecular line observations toward IRAS~23304+6147 at the 1.3\,mm band were performed using  the Northern Extended Millimeter Array. 
A morpho-kinematic model was built to reproduce the observed
$^{13}$CO images and to decipher the structures of the nebula.
}
 {The imaging results reveal an elliptically elongated shell 
with an equatorial density enhancement (or a torus-like structure),
 and in detail how the various molecules distribute in the envelope. 
 The nebular morphology  points to a binary system in which the ultraviolet radiation from the companion may trigger photochemistry in the inner regions.
 The torus-like structure exhibits an enrichment of linear carbon-chain molecules and a depletion of silicon-bearing molecules. 
 This PPN has a mass lower limit of $1.3\times10^{-2}$\,M$_\odot$ 
 and may exhibit a low $^{12}$CO/$^{13}$CO abundance ratio.}
{The chemically stratified structure of 
$^{13}$CN, HC$_3$N, and C$_4$H
represents an observational evidence
of the internal radiation that initiates photochemistry.
 The carbon-rich torus-like structure probably offers
a conducive environment for the formation of dust and complex molecules implicated in the rare 21\,$\mu$m emission.  We hypothesize that the  21\,$\mu$m 
sources probably descend from J-type carbon stars
via a binary evolutionary channel.
}


\maketitle
%

\section{Introduction}\label{intro}
When low- and intermediate-mass stars evolve into the 
asymptotic giant branch (AGB) stage, intense stellar winds emanate from the stellar surface, propelling the circumstellar material into interstellar space at mass-loss rates typically ranging from $10^{-8}$--$10^{-5}$\,M$_\sun$~yr$^{-1}$ and reaching as high as $10^{-4}$\,M$_\sun$~yr$^{-1}$
\citep[e.g.][]{1993ApJ...413..641V,1995A&A...299..755B,2019NatAs...3..408D}. 
The highly opaque AGB envelope is an enormous chemical
factory \citep[e.g.,][]{cer00}, where over 100 gas-phase molecules and 15 solid-state species have been detected \citep{2021ARA&A..59..337D}.
After the high-mass loss ceases, the circumstellar material is gradually detached from the photosphere and the stellar temperature increases. Before the stellar ultraviolet (UV) radiation becomes intense enough to ionize the envelope, there is a brief post-AGB phase, during which the star is surrounded by a protoplanetary nebula (PPN). The PPN stage lasts approximately $10^3$ years. In contrast to the AGB envelope, whose morphology is predominantly spherical, the PPN's shape starts to deviate from 
overall spherical symmetry, becoming more complex \citep[e.g.,][]{1998AJ....116.1357S}.

Recently, high angular resolution observations have
revealed various structures in  AGB envelopes,
which may be attributed to the interaction of a companion
\citep{2012Sci...338..773B,2020Sci...369.1497D}. 
The binary system plays an important role in the shaping
of PPNe during the transition from AGB to post-AGB
\citep[see][for a review]{2018Galax...6...99L}.

Apart from the morphology transition, PPNe show a chemical complexity.
It is well established that circumstellar chemistry is determined by the C/O abundance ratio as the less abundant element is mostly locked up in CO. Carbon-rich circumstellar envelopes are dominated by carbonaceous molecules such as , such as CS, CN, HCN, and HC$_3$N \citep[e.g.,][]{1993ApJS...87..267O}, 
whereas oxygen-rich envelopes predominantly contain oxygenated species like SO, OH, SO$_2$, and H$_2$O \citep[e.g.,][]{1979A&AS...36..337E}.
 A large variety of gas-phase molecules have been
detected in the prototype carbon-rich PPNe
CRL\,2688 and CRL\,618 through spectral line surveys
with single-dish telescopes \citep{2007ApJ...661..250P,2008AJ....136.2350P,2013ApJ...773...71Z,2022ApJS..259...56Q}.
There is ample observational evidence showing that aromatic and aliphatic materials as well as C$_{60}$ can be quickly synthesized in the PPN phase \citep{2001ApJ...554L..87K,2011ApJ...730..126Z}. 

A quintessential example of the chemical complexity in PPNe is the discovery of the 21\,$\mu$m feature.
This feature was discovered for the first time by \citet{1989ApJ...345L..51K}, and has been detected so far in only 31 objects \citep[20 in the Galaxy, 9 in the Large Magellanic Cloud, and 2 in the Small Magellanic Cloud; see][for a recent review]{2020Ap&SS.365...88V}. All the 21\,$\mu$m  sources are
carbon-rich PPNe with largely enriched $s$-process elements
\citep{2003ARA&A..41..391V}. The rare detection 
indicates a transient nature.
Observations show that the 21\,$\mu$m feature is always accompanied by a broad feature at 30\,$\mu$m, with the former accounting for about 1$\%$--5$\%$ of the infrared emission of the PPN, and the latter accounting for about 7$\%$--20$\%$ \citep{2010ApJ...725..990Z}; but there is no correlation between the intensities of the two features \citep{2010EP&S...62..105J}.
The 21\,$\mu$m feature remains unidentified although a dozen different species, ranging from individual carbon- or silicon-bearing molecules to polycyclic aromatic hydrocarbon  complexes, have been proposed as the potential carriers
\citep[see the references in][]{2020Ap&SS.365...88V}.
Some of them can be ruled out because of the insufficient cosmic abundance of the incorporated elements or the non-detection
of sub-features arising from the same species
\citep{2009MNRAS.396.1247Z}.
It is unclear why this feature abruptly appears and rapidly disappears and what role its carrier could play on circumstellar chemistry.
Unraveling its origin is crucial to deepening our understanding of the enrichment of the interstellar medium.

The exclusive presence of the 21\,$\mu$m feature
in carbon-rich PPN suggests a significant link to carbon-chain chemistry. It follows that investigating carbon-bearing  molecules within 21\,$\mu$m sources could reveal critical insights into the identification of this mysterious feature \citep{2020ApJ...898..151Z,2024AJ....167...91Q}. However, the scarcity of 21\,$\mu$m sources, the inherent faintness of molecular lines in these sources, and our incomplete understanding of circumstellar chemistry during the brief PPN period present significant challenges to this research.  As a crucial first step of investigating the physicochemical environments necessary for the formation of the 21\,$\mu$m carrier, we need to perform high-sensitivity observations and  statistically study the nebular characteristics that are common for the 21\,$\mu$m sources. Previous efforts have been mostly based on single-dish observations although interferometric observations of CO emission in a few 21\,$\mu$m sources have been presented \citep{2009ApJ...692..402N,2012ApJ...759...61N,2019IAUS..343..520U},  which are unable to reveal the chemical complexity induced by the morphology evolution of PPNe.

In this paper, we present millimeter interferometry observations of
a prototypical 21\,$\mu$m source IRAS~23304+6147, aiming
to investigate the nebular structure and the spatial distributions of the gas-phase molecules.
The observations and data reduction are described in 
Section~2. 
The spectra and images of the identified molecules are presented in Section 3, along with a morpho-kinematic model that is developed to explain the data.
In Section~4 we discuss the nature and origin of the nebular structure, the $^{12}$CO/$^{13}$CO abundance ratio, and the implications on circumstellar chemistry.
Our conclusions follow in Section~5.


\section{Observation and Data Reduction} \label{sec:obse}

IRAS~23304+6147 was observed utilizing the Northern Extended Millimeter Array (NOEMA) Interferometer
 in configuration 12A on 2022 February 23 (1.9~h on source),  February 28 (2.9~h on source), and  March 6 (0.8~h on source) under the project number W21BI (PI: Y. Zhang). 
The precipitable water vapor during the
observations spreads over 0.4--1.2\,mm. The system temperatures range from 100--150\,K.
Twelve antennas were using with  baselines ranging from 32--920\,m.
The observations were performed with the lower sideband (LSB)  covering 213--221 GHz and the upper sideband (USB) covering 228--236 GHz. Doppler tracking was employed, establishing a central velocity of
$-16.1$\,km~s$^{-1}$, the systematic velocity of IRAS~23304+6147,
relative to the local standard of rest (LSR). The phase calibrators are  0059+581, J0011+707, and J2250+558; the bandpass calibrators are 0059+581 and 3C84. 
MWC349 and 2010+723 were observed for flux calibration, which
have  flux densities of 2.1 and 0.6\,Jy, respectively.

The phase center was set at $\left(\alpha_{2000}, \delta_{2000}\right)=\left(23^{\mathrm{h}} 32^{\mathrm{m}} 44^{\mathrm{s}} 785, +62^{\circ} 03^{\prime} 49.084^{\prime \prime} \right)$.
Data were calibrated  using the \texttt{GILDAS/CLIC} 
package\footnote{\url{http://www.iram.fr/IRAMFR/GILDAS}}. 
Imaging and deconvolution processes were conducted using \texttt{GILDAS/MAPPING} and \texttt{GILDAS/IMAGER}\footnote{\url{https://imager.oasu.u-bordeaux.fr}}.
The CLEAN algorithm of \citet{1974A&AS...15..417H}  was employed
in the imaging process.
Uniform weighting was used in order to optimize
the spatial resolution, resulting in a synthesized beam of $0.52\arcsec \times 0.42\arcsec$ at a position angle (PA) of $23^\circ$.  Molecular maps were obtained by subtracting
the maps of continuum emission  extracted from line-free spectral channels. The Root-mean-square (rms) noise of the line channel maps
 is 1.25--1.61 $\rm mJy~beam^{-1}$. The spectral resolution is about 2~MHz (2.7\,km~s$^{-1}$) at 1.3~mm.

\section{Results} \label{sec:style}

\subsection{Spectral lines}

\begin{figure*}
   \centering
\includegraphics[width=0.9\textwidth]{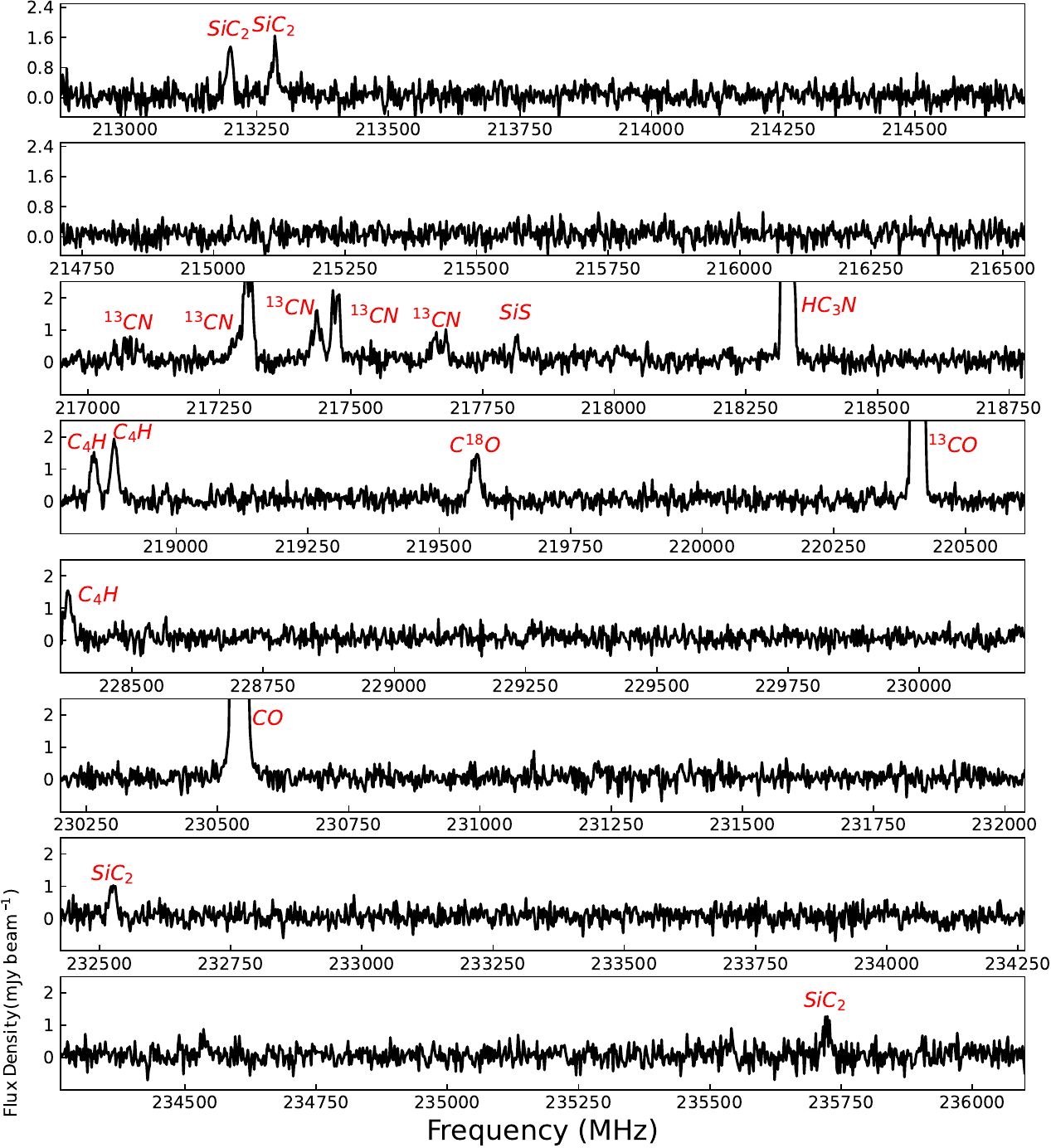}
   \caption{NOEMA spectra of IRAS~23304+6147 in the frequency ranges 
   of 213–221 GHz (LSB) and 228–236 GHz (USB).}
   \label{spec}
\end{figure*}

\begin{figure*}
   \centering
\includegraphics[width=1\textwidth]{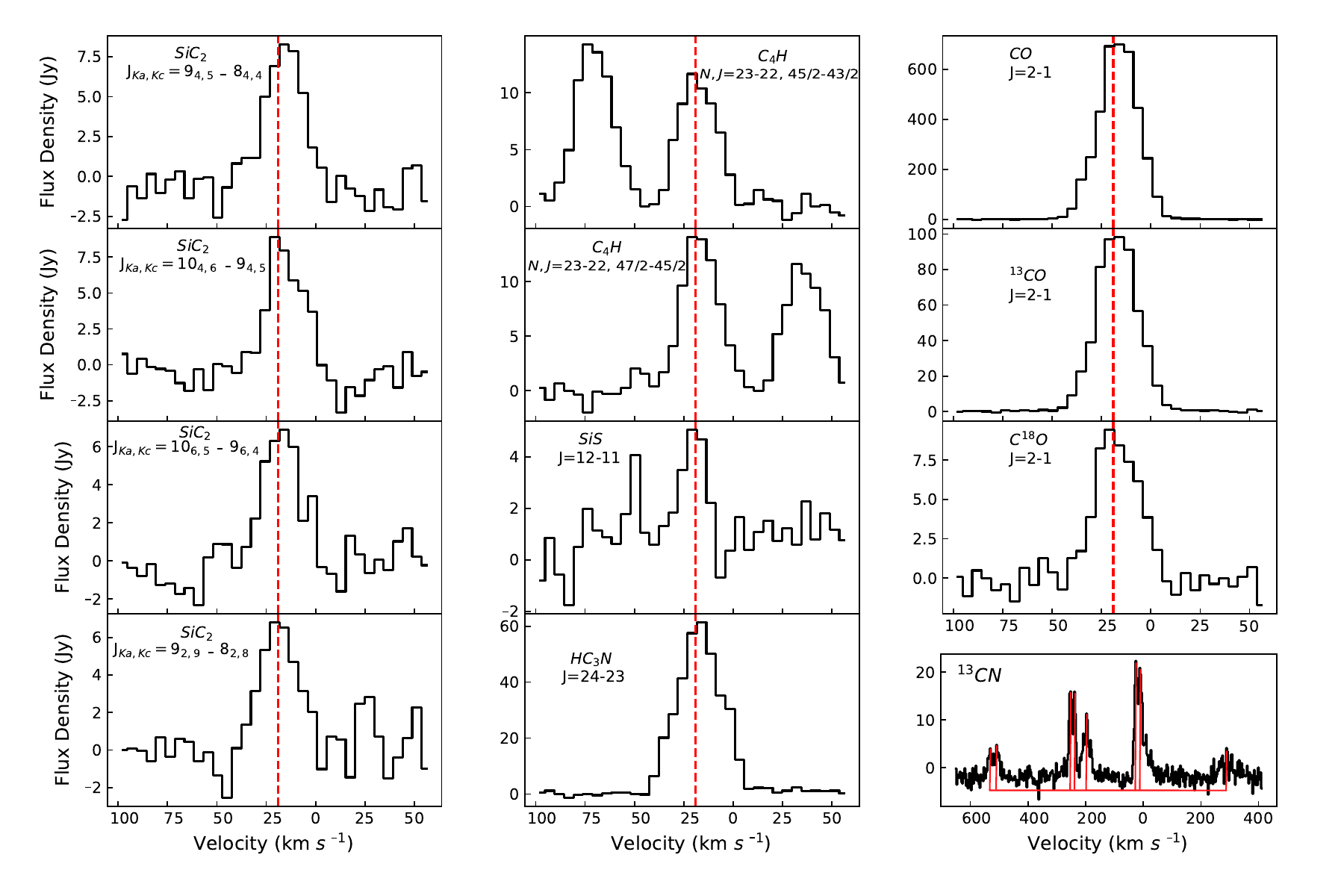}
   \caption{Profiles of the lines detected in
 IRAS~23304+6147. 
 The systemic velocity ($-16.1$\,km~s$^{-1}$) is indicated by the vertical dashed line. The positions and   
   relative intensities of the hyperfine components
   of  $^{13}$CN are marked  by vertical solid lines.}
   \label{spec_detail}
\end{figure*}

Previously reported single-dish observations toward IRAS~23304+6147 have detected CO ($J=1$--0 and 2--1) \citep{1990A&A...228..503W,1991A&A...246..153L,2005ApJ...624..331H}, $^{13}$CO ($J=1$--0 and 2--1) \citep{2012ApJS..203...16S, 2024AJ....167...91Q}, CS ($J=3$--2) \citep{1993A&A...267..515O}, HCN ($J=1$--0 and 3--2), H$^{13}$CN ($J=3$--2), and  CN ($N=2$--1) \citep{2024AJ....167...91Q}. Our interferometric observations result in detections of 17 individual emission features belonging to eight different molecules and isotolopogues, among which  C$^{18}$O, $^{13}$CN,  HC$_3$N, C$_4$H, SiS, and SiC$_2$ are observed for the first time in this source. Figure~\ref{spec} shows the spectrum obtained by averaging the flux densities over a $2\arcsec\times2\arcsec$ area centered on the phase center. Line measurements are presented in Table~\ref{line}, where E$_u$ is the energy of the upper level, and  $I_{\rm peak}$ and $\int I \, dv$ represent the peak and integrated intensities, respectively.
The line profiles are shown in Figure~\ref{spec_detail}, where the flux density has been
 converted to Jy using a conversion factor of approximately 32.8 Jy/K.

\begin{figure*}
   \centering
\includegraphics[width=0.8\textwidth]{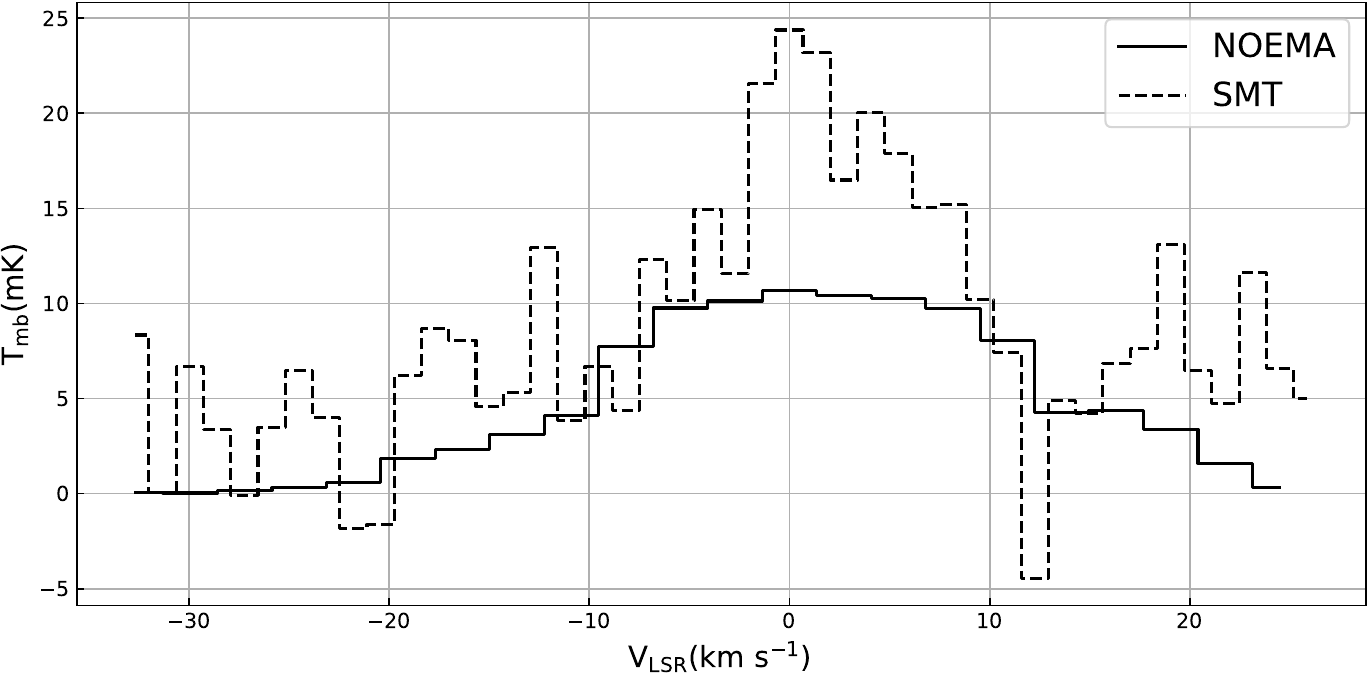}
   \caption{Comparison of the $^{13}$CO $J=2$--1 observations made by NOEMA and SMT. The NOEMA data have been convolved to match the SMT's resolution
   of 34.3\arcsec.  The central velocity is set to 0\,km~s$^{-1}$.}
   \label{compare_smt_noema}
\end{figure*}

In order to estimate the flux loss of our NOEMA observations,  
we compare the single-dish data of the
CO and $^{13}$CO ($J=2$--1) lines observed by
the Heinrich Hertz Sub-millimeter Telescope (HHT) and
the 10m Arizona Radio Observatory Submillimeter Telescope (SMT),
respectively \citep{2005ApJ...624..331H,2024AJ....167...91Q}. 
Figure~\ref{compare_smt_noema} shows a comparison of the $^{13}$CO ($J=2$--1) observations made by NOEMA and SMT. 
According to the single-dish observations, 
the CO and $^{13}$CO lines have peak intensities of
0.2 and 0.025\,K and integrated intensities of 3
and 0.33\,K\,km\,s$^{-1}$, respectively.
The half-power beamwidths (HPBWs) of HHT and SMT 
are 32$\arcsec$ at 230\,GHz and 34.3$\arcsec$ at 220\,GHz,
respectively. According to the NOEMA images,
the source size is roughly $2\arcsec$, resulting in filling factors of 0.0039 and 0.0034 for the HHT and SMT beams, respectively.
If there is no flux loss, 
the CO and $^{13}$CO lines should have peak intensities of 90 and 12\,mK and integrated intensities of 0.93 and 0.12\,K\,km\,s$^{-1}$ in the HHT and SMT observations, respectively. This indicates a flux loss of $70\%$ in our NOEMA CO and $^{13}$CO line observations. 
Extended fossil AGB halos have been revealed in other PPNe \citep[e.g.,][]{2007AJ....134.2200S}.
It is likely that an AGB halo has been
substantially filtered out in our observations.
Presumably, the flux loss is  less severe for other lines that have a spatial extension significantly smaller than CO.
 
\subsection{Maps}

\begin{figure*}
   \centering
\includegraphics[width=0.9\textwidth]{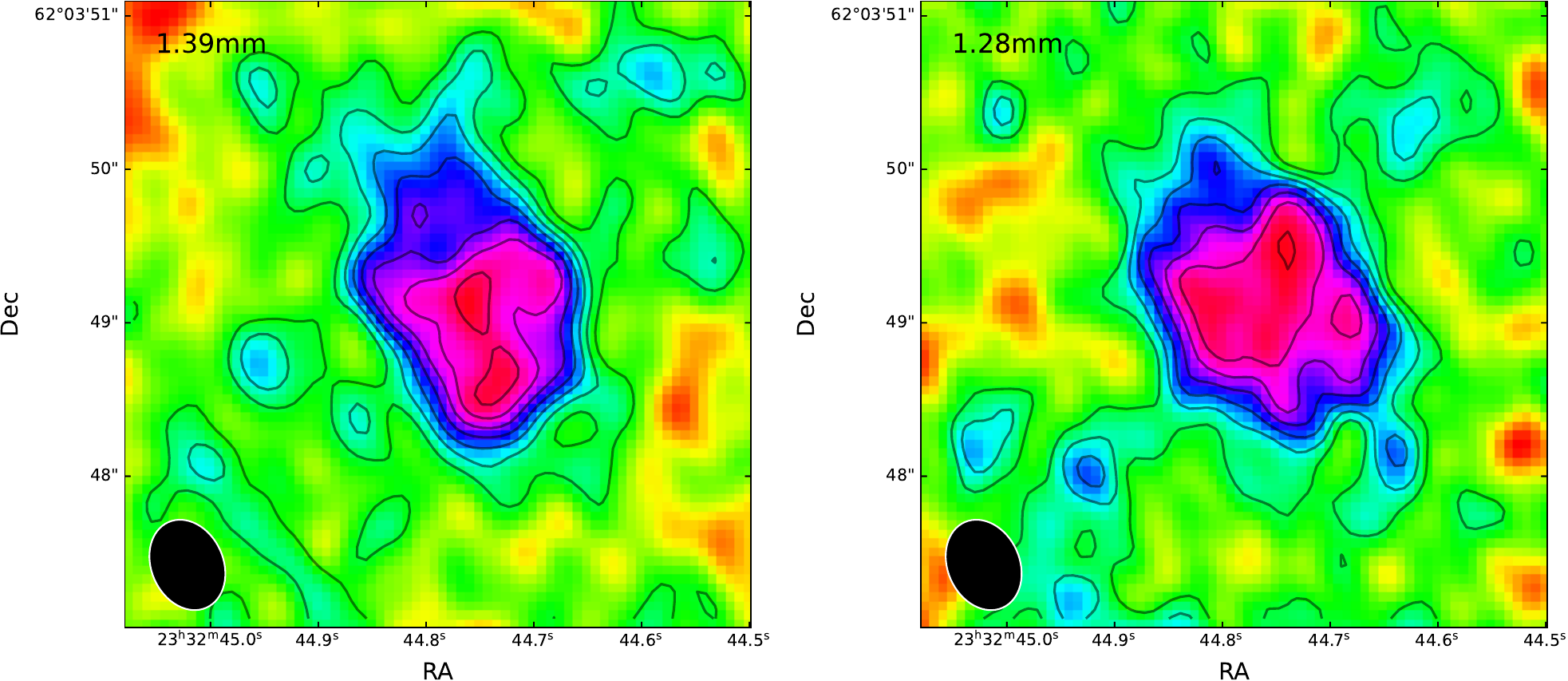}
   \caption{Continuum emission at 1.39 and 1.28\,mm wavelengths. The synthesized beams
   shown in the down-left corner are of size
   $0.52\arcsec\times0.42\arcsec$. 
   Contour levels are  at 1$\sigma$ intervals starting at 1$\sigma$, where the rms $\sigma=41$  and 55\,$\mu$Jy $\rm beam^{-1}$ at 1.39 and 1.28 mm, respectively.}
   \label{cont}
\end{figure*}

Figure~\ref{cont} shows the maps of continuum emission 
at 1.39 and 1.28\,mm, which have flux densities
of 3.67 and 3.57\,mJy, respectively, indicating an in-band spectral index\footnote{The spectral index $\alpha$
is defined as
the exponent in the power-law relation describing the dependence of the continuum flux density $S_\nu$ on frequency $\nu$,
given by $S_\nu\propto\nu^\alpha$.} 
of $\alpha=-0.3$. 
  The derived $\alpha$ value
appears to be in close proximity to the theoretical predictions of
 optically thin free-free emission ($-0.1$).
However, since the central star of this PPN lacks sufficient ionizing flux to sustain extended ionized regions, the physical
origin of the spectral index remains unclear.
Owing to the restricted frequency range and feeble continuum emission, we exercise caution and refrain from over-interpreting the result. 

Figure~\ref{compare_molecule} shows the continuum-subtracted
moment 0 maps of molecular lines.
Overall, the post-AGB nebula appears as an elongated ellipsoid 
(EE) with an equatorial density enhancement (EDE). 
The presence of the EE structure is consistent with previous 
optical observations \citep{2007AJ....134.2200S} and is typical of 
 nascent PPNe \citep{1992A&A...257..701B}.
The NOEMA maps show that different molecules have notably different spatial distributions,
demonstrating a significant chemical stratification. 
The optically thick CO emission shows an oval-shaped structure
with an inner bright ring and an outer halo. The EDE structure
is more striking in the $^{13}$CO map, and manifest itself
as  a torus-like structure in the maps of 
linear carbon-chain molecules (C$_4$H and HC$_3$N).
It should be noted that the gaseous torus-like structure has also been detected in other 21\,$\mu$m sources
IRAS~07134+1005, IRAS~22272+5435, and IRAS~16594$-$4656
\citep{2009ApJ...692..402N,2012ApJ...759...61N,2019IAUS..343..520U}.
Despite their very modest emission,  silicon-bearing molecules (SiS and SiC$_2$) seem to be distributed in the peripheral regions of the envelope with fragmentary structures.
Intriguingly, the SiS and SiC$_2$ fragments do not coincide in
position. 
The $^{13}$CN emission is dominantly concentrated in the center,
while the C$^{18}$O emission is too faint to draw any conclusion
about its spatial distribution.

\begin{figure*}
    \centering
\includegraphics[width=1\linewidth]{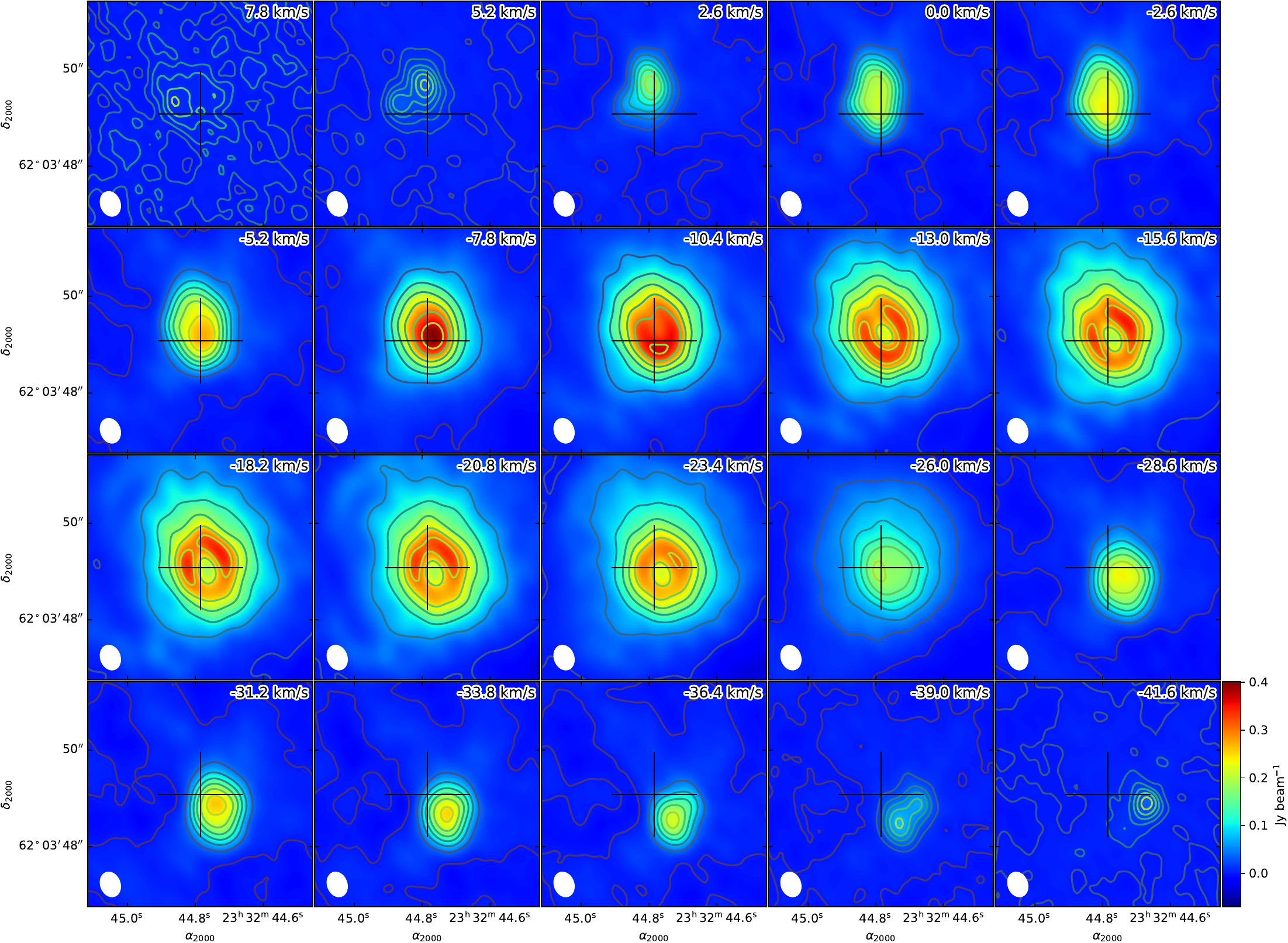}
    \caption{Channel maps of IRAS~23304+6147 in 
    CO $J=2-1$ emission. The contour levels are individually set for each channel by creating ten equally spaced contour lines between the maximum and minimum intensities of that channel. The crosses mark the position of the phase center.
     The LSR velocities are indicated in each panel. 
     The systemic velocity of this object is $-16.1$\,$\rm km~s^{-1}$.
     The white ellipse at the bottom-left corner represents the synthesized beam, with a size of $0.52\arcsec \times 0.42\arcsec$  and a PA of $23^\circ$. }
    \label{co_channel}
\end{figure*}

\begin{figure*}
    \centering
\includegraphics[width=1\linewidth]{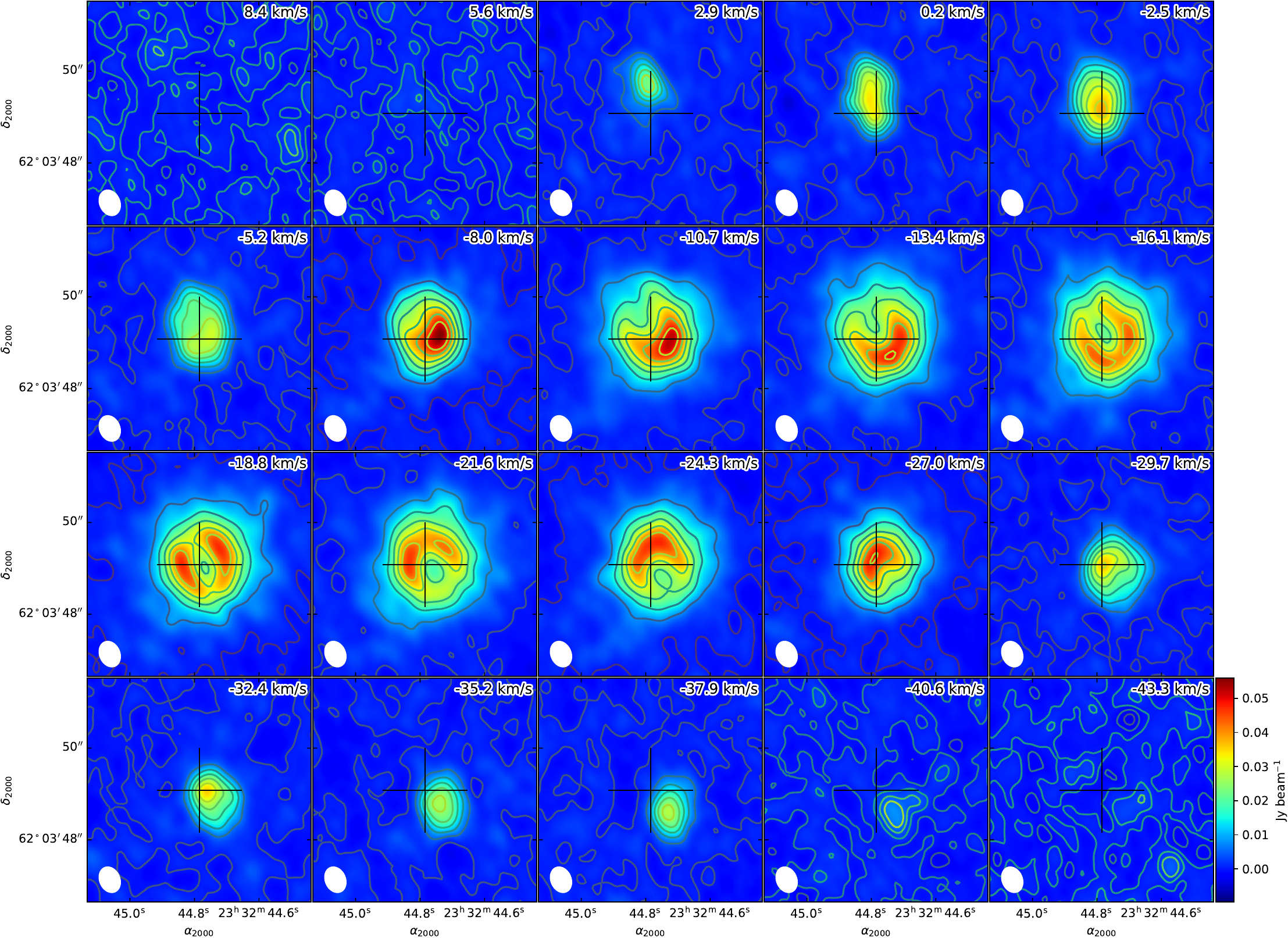}
    \caption{Same as in Figure~\ref{co_channel}
    but for  $^{13}$CO $J=2-1$.}
    \label{13co_channel}
\end{figure*}

\begin{figure*}
    \centering
\includegraphics[width=1\linewidth]{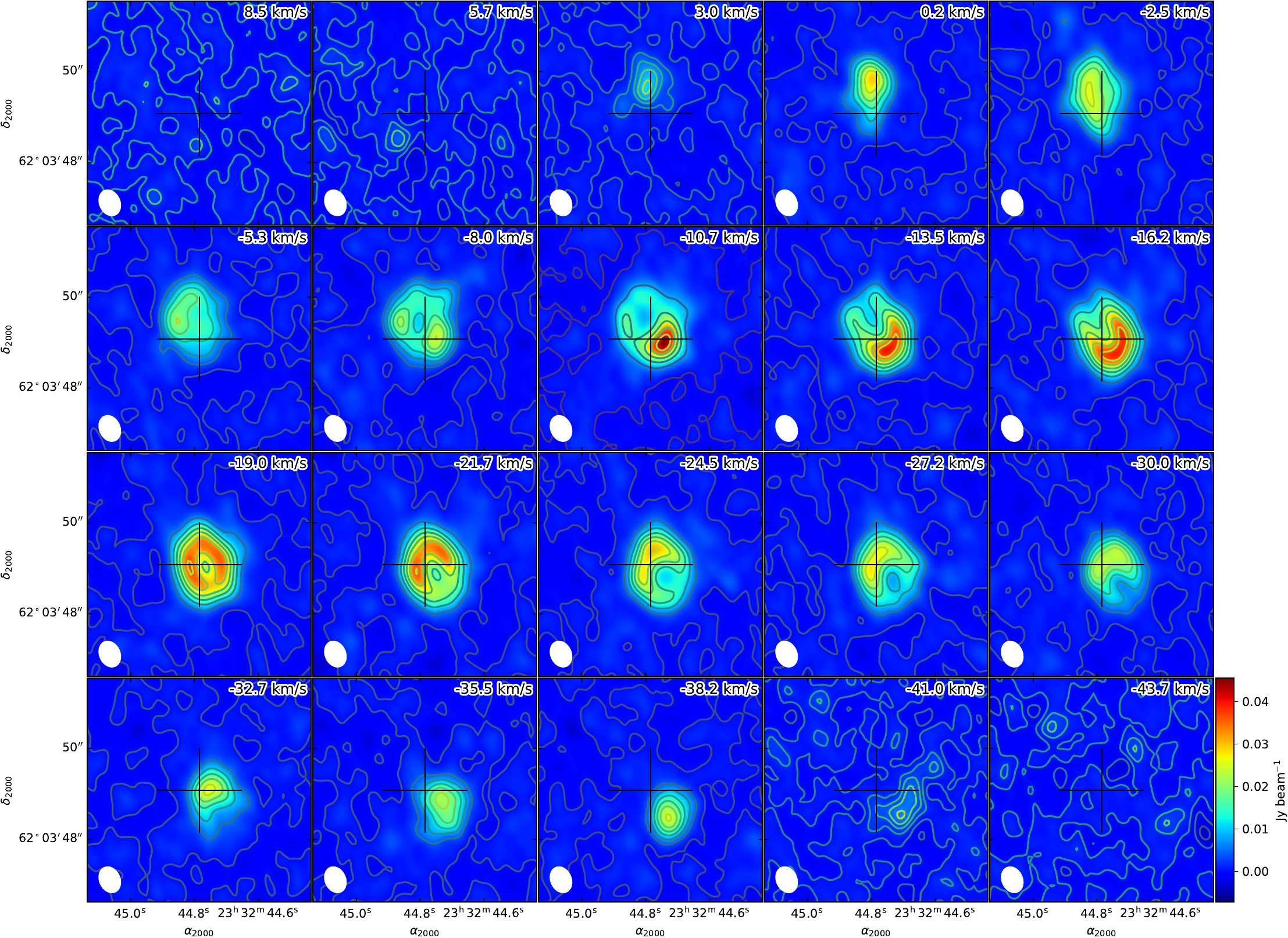}
    \caption{Same as in Figure~\ref{co_channel}
    but for  HC$_3$N $J=24-23$.}  
    \label{hc3n_channel}
\end{figure*}

The channel maps of CO, $^{13}$CO, and HC$_3$N lines are shows in Figures~\ref{co_channel}--\ref{hc3n_channel}. An inspection of these maps reveals that the intensity distributions highly depend on the velocities.
In the red channels ($V_{\rm LSR}>-3$\,km~s$^{-1}$)
the EE appears as a bright spot located in the northwest 
region, while in the blue channels
 ($V_{\rm LSR}<-29$\,km~s$^{-1}$) the bright region of
EE is shifted to the southeast. 
The behavior of the EDE in the channel maps is opposite to that of the EE. 
As seen in the channel maps of
$^{13}$CO and HC$_3$N (Figures~\ref{13co_channel} and \ref{hc3n_channel}),
the crescent-shaped EDE emission
is shifted from the southeast to the northwest from the red to the blue
channels ($V_{\rm LSR}=-8$ to $-27$\,km~s$^{-1}$).
This is a clear demonstration showing that the major axis of the EE is inclined toward us in the southeast, and the EE expands
along the major axis faster than the EDE\footnote{The EE expands at different
velocities along different directions, and has the highest
velocity along the major axis, which is referred to the velocity
of the EE hereafter for simplicity. The EDE expands along
the equatorial direction at a constant velocity.}.

\subsection{Morpho-kinematic model}

To investigate the structures of IRAS~23304+6147, we employ the three-dimensional morpho-kinematic modelling code SHAPE \citep{2012ascl.soft04010S}. 
We first input the initial parameters into SHAPE. Subsequently, through an iterative procedure, we generate synthetic images as well as position-velocity (P-V) diagrams that closely resemble the observational data. From the free parameters refined during this iterative process, we proceed to deduce the morpho-kinematic structure of the model. This methodology has been previously demonstrated to be successful in reconstructing PPNe in other research works
\citep[e.g.,][]{2011ApJ...740...27K,2012ApJ...759...61N,2015A&A...573A..56S,2019A&A...629A...8T}.

\begin{figure*}
    \centering
    \begin{minipage}[t]{0.3\linewidth}
        \centering
        \includegraphics[width=\linewidth]{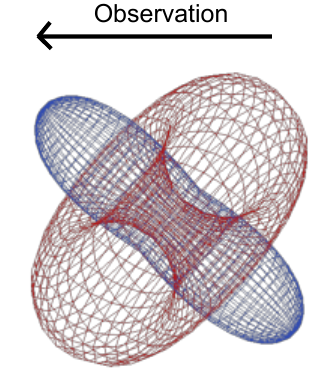}
    \end{minipage}
    \hspace{0.07\linewidth}
    \begin{minipage}[t]{0.3\linewidth}
        \centering
        \includegraphics[width=\linewidth]{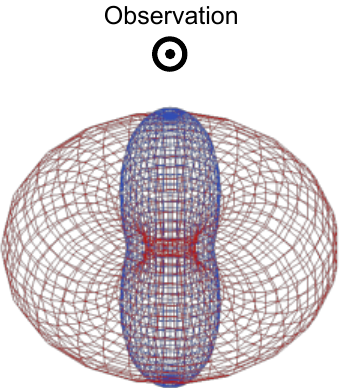}
        \label{face_model}
    \end{minipage}
    \caption{
    Three-dimensional mesh geometry of the SHAPE model, as
 seen from two orthogonal directions.  The left panel shows the side view along the observer’s direction, while the right panel shows the front view facing the observer. The red torus represents the EDE, and the blue peanut-shaped structure represents the EE. The inner and outer radii of the EDE are assumed to be 0.78 and 1.75 \arcsec, respectively. The major axis size of the EE is assumed to be 2 \arcsec.}
       \label{shape_model}
\end{figure*}

The model consists of two
components corresponding to the EDE and EE, as shown in 
Figure~\ref{shape_model}. We assume that the EDE and the EE are symmetric in the azimuthal direction around the EE's major axis. 
In a cylindrical coordinate system, we set the major axis of the EE as $z$-axis. Then, we can describe the molecular density distributions using a formula with the variables $z$ and $r$, where $z$ represents
the vertical displacement relative to the central star, and $r$ represents the orthogonal distance to the $z$-axis.
We assume that the molecular gas is confined within 
the meshes shown in Figure~\ref{shape_model} and
the densities of the EDE and the EE follow the formulas,
\begin{equation}
n_{EDE}(r)=n_1 e^{-\frac{\left(r- r_1\right)^2}{2 \sigma_{1}^2}},
\label{density_ede} 
\end{equation}
and
\begin{equation}
n_{EE}(z)=n_2 e^{-\frac{z^2}{2 {\sigma_2}^2}}
\label{density_ee}, 
\end{equation}
A simple Hubble flow is adopted for the velocity fields of the EDE and EE.
Specifically, the maxima velocity of the EE is 2.5 times that of
the EDE.

\begin{figure*}
    \centering
\includegraphics[width=1\linewidth]{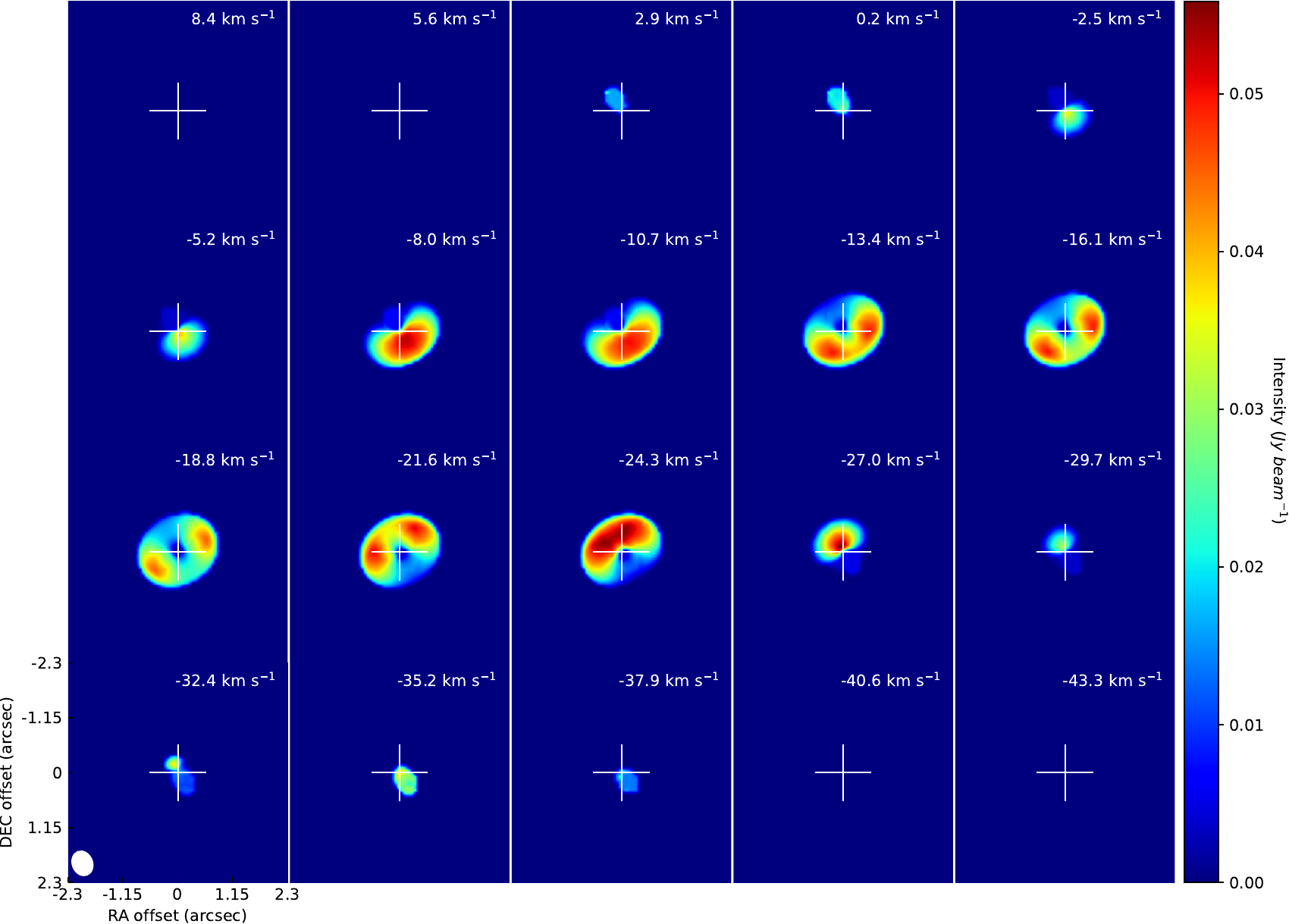}
    \caption{Modeled $^{13}$CO ($J=2$--1) channel maps.
 The white ellipse in the lower-left corner of the image represents the synthesized beam.  The angular size of each panel is identical to
 those in Figures~\ref{co_channel}--\ref{hc3n_channel}.
 }  
    \label{channel_model}
\end{figure*}

\begin{figure*}
    \centering
\includegraphics[width=1\linewidth]{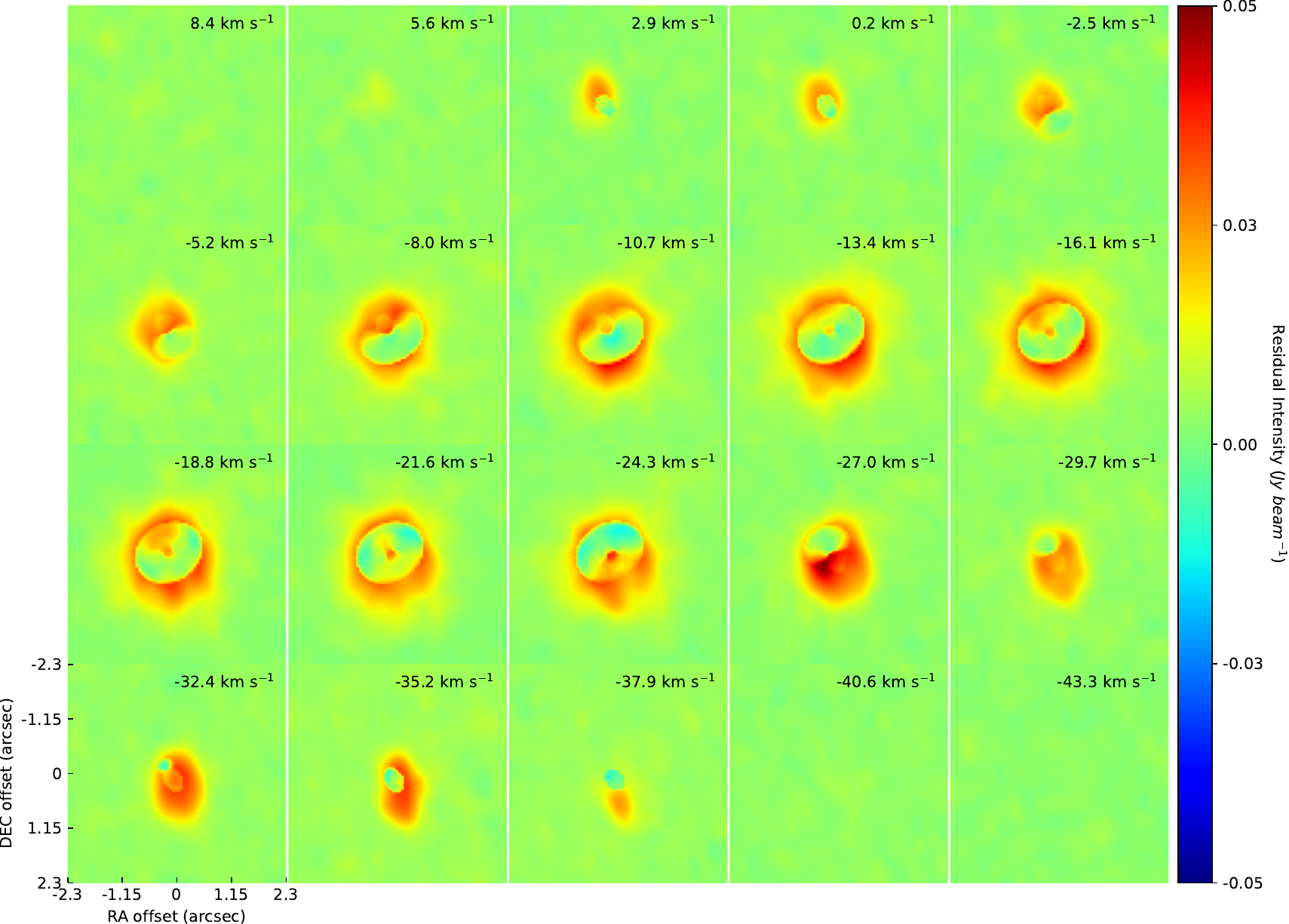}
    \caption{Residuals between the observed and modeled $^{13}$CO ($J=2$--1) channel maps. The angular size of each panel is identical to those in Figures~\ref{co_channel}--\ref{hc3n_channel}.}
    \label{residual_map}
\end{figure*}

Given the assumption that the $^{13}$CO ($J=2$--1) line
is optically thin, which implies that its surface brightness is
proportional to the $^{13}$CO column density, we utilize
the $^{13}$CO channel maps  to impose constraints on the density distribution and velocity fields within the model\footnote{The $^{13}$CO ($J=2$--1) line may not be optically thin in PPNe. For instance, 
\citet{2013A&A...557L..11B} show that  the Red Rectangle's core is optically thick in $^{13}$CO ($J=2$--1) line. However,
the morpho-kinematic reconstruction remains robust to 
slight deviations from the optically thin assumption.}.
Figures~\ref{channel_model} and \ref{residual_map}  present the 
modeled channel maps and the residual maps, respectively.
The optimized modelling parameters in Equation~\ref{density_ede} and Equation~\ref{density_ee} are as follows: 
$n_1=1.9\times10^{7}$~cm$^{-3}$,
$r_1=2.8\times10^{16}$\,cm, $\sigma_1=4.6\times10^{15}$\,cm, $n_2=5.4\times10^{6}$~cm$^{-3}$,
and $\sigma_2=2.2\times10^{16}$\,cm.

The model successfully reproduces the main observable features of the primary nebula on the channel maps although it cannot  precisely replicate the surface brightness. The existence of the EDE-EE structure is unequivocally corroborated, and the modeled velocities of the EDE and EE should be reliable. The model's performance deteriorates at the tip of the EE. In comparison to the observations, there is an  displacement, which 
can be partially attributed to the oversimplified assumption of the
density distribution and the photodissociation of $^{13}$CO (see Section~\ref{dis423}). 
Furthermore, as illustrated in Figure~\ref{residual_map}, approximately 20{\%} of the total flux arises from a diffuse  component not included in our current model. 
The diffuse emission exhibits remarkable intensity in the regions adjacent to the EDE and can be seen in all of the velocity channels where the EDE and EE are present. This extended emission can be attributed either to part of an AGB halo expanding at 
 $\sim20$\,km\,s$^{-1}$ or a disk-wind associated with the EDE.

Figure~\ref{pv_model} shows a comparison between the observed P-V diagram and the modelling results. The kinetic structures of the  EE and EDE are well traced by the P-V diagrams of $^{13}$CO. We do not see signature of rotating disk that has been detected in a handful of PPNe \citep{2021A&A...648A..93G}. The deduced inclination angles from the sky plane of the EDE and the EE are 35$^\circ$ and 55$^\circ$, respectively. According to the modelling, the EE and EDE have maxima expansion velocities of about 35 and 14\,km\,s$^{-1}$, respectively. The primary axis of the EE has an angular dimension of 2$\arcsec$, whereas the inner and outer diameters of the EDE are 0.78$\arcsec$ and 1.75$\arcsec$. Adopting a distance of 4.32\,kpc \citep{2022ApJ...927L..13K}, we estimate the inner and outer diameters of the EDE to be 3300 and 7500\,AU and the EE primary axis to be 8600\,AU in size. This results in dynamic ages of 340--770 yr and 600 yr for the EDE and EE, respectively.  Consequently, we infer that the EE and EDE were formed approximately at the same time. 

Although the values of the parameters used in the model are carefully chosen based on the observational data and our current understanding of stellar winds,
we give no warranty that they represent the only possible solution. It is crucial to point out that the SHAPE modelling is intended to examine the morpho-kinematic structures of the nebula on relative scales. We have not carried out radiative transfer calculations in this model. Consequently, the results do not directly hinge on the temperature. 
In the future, it is highly desirable to develop a sophisticated radiative transfer model that incorporates $^{12}$CO lines, which is more sensitive to the temperature, in order to achieve more precise results.

\subsection{Nebular Mass}

Once the excitation temperature is established, the mass of molecular gas ($M$) can be derived from the  
$^{13}$CO ($J=2$--1) line through the formula given
by \citet{1999A&A...347..194O} and \citet{2006ApJ...645..605C}.
 Drawing upon the approach delineated by \citet{2009ApJ...692.1382D}, we estimate the kinetic temperature (\(T_K\)) of the molecular gas to be approximately 10\,K. Subsequently, we made use of the online radiative transfer modelling tool RADEX \citep{2007A&A...468..627V} to compute the excitation temperature of the  $^{13}$CO ($J=2$--1) transition, which is roughly 10\,K. Based on the same assumption for IRAS~08544\(-\)4431 \citep{2009ApJ...692.1382D}, where \(f_{\rm 13CO} = 2\times10^{-5}\), we find \(M = 1.3\times10^{-2}\,\mathrm{M_\odot}\) for the nebula around the post-AGB star IRAS~23304+6147. There are some uncertainties in calculating \(M\). Regarding IRAS~08544\(-\)4431, the investigations carried out by \citet{2009ApJ...692.1382D} and \citet{2018A&A...614A..58B} derived values of \(M = 4.7\times10^{-3}\,\mathrm{M_\odot}\) and \(2.2\times10^{-2}\,\mathrm{M_\odot}\), respectively. 
  After factoring in the effect of the different distances used in these calculations, this discrepancy indicates a five- to 
 sixfold uncertainty in the mass estimation. 
 \citet{2018A&A...614A..58B} implied a value of \(f_{\rm 13CO} = 1.5\times10^{-5}\) for IRAS~08544\(-\)4431, which is comparable to the assumption made by \citet{2009ApJ...692.1382D}. However, their studies indicated that the disk temperature in IRAS~08544\(-\)4431 could potentially exceed 100\,K. If we assume \(T_K = 100\,\mathrm{K}\), \(M\) will decrease by a factor of 1.5. Therefore, we estimate that the uncertainty in the calculated mass could be as large as a factor of eight. Because of flux loss, the derived \(M\) should be regarded as a lower limit to the nebular mass. There might be an extended AGB-remnant halo with a mass of \(3\times10^{-2}\,\mathrm{M_\odot}\).

\section{Discussion} \label{sec:floats}

The NOEMA observations of IRAS~23304+6147 update the molecular richness of this source and offer new details about its nebular structure that was not disclosed by earlier optical measurements.
In Section~\ref{disa}, we discuss the shaping of this PPN.
In Section~\ref{disb}, we examine the connection between 
circumstellar chemistry and nebular structure, which may shed new light on the origin of the 21\,$\mu$m feature.

\subsection{Origin and nature of the EDE}\label{disa}

As pointed out by \citet{2018Galax...6...99L}, there are
two types of EDEs. 
The first type is short-lived, created prior to the jet, and has a slow expansion velocity and limited angular momentum.  
One such example is that found in the water fountain  IRAS~16342-3814 
\citep{2017ApJ...835L..13S}. The second type
is in Keplerian rotation with a large angular momentum,
and has a longer lifetime. Disks of this kind commonly coexist with disk-winds, as demonstrated by observations of the PPNe AC Her and Red Rectangle \citep{2015A&A...575L...7B, 2016A&A...593A..92B}. 
Observational studies \citep{2021A&A...648A..93G,2023arXiv231108034G} 
have demonstrated that although these outflows generally exhibit lower mass 
budgets compared to their progenitor disks, 
they can occasionally exceed the mass of
the rotating components they surround.
 Apart from the disk-winds, high-velocity jets may be generated from the companion star \citep{2022A&A...666A..40B}.
 In contrast to disk-winds, these collimated high-velocity jets make only a negligible contribution to the overall nebular mass
  \citep{2023arXiv231108034G}.
As shown in Figure~\ref{pv_model}, the EDE of IRAS~23304+6147  has a slow expansion velocity of $\sim20$\,km\,s$^{-1}$.
For a Keperian disk, the velocity should be inversely proportional to the radial distance from the center, however this is not 
the case in Figure~\ref{pv_model}. Therefore,
the EDE does not exhibit rotation, and should be assigned in the first category. The formation of EDE is commonly attributed to the interaction in a binary system, where  the mass loss from the primary star tends to be directed in the orbital plane in the gravitational field of the companion. A pre-existent circumbinary disk may slow down the 
super stellar winds along the equatorial direction, generating the EDE. Dynamic interaction could lead to an angular momentum transfer from the binary orbit to the EDE, forming an expanding torus structure.  During further evolution the torus may feed a circumpanion accretion disk, leading to the launch of a jet \citep[see][and reference therein]{2022A&A...666A..40B}.

\citet{2020Sci...369.1497D} show that in a binary system the favourable conditions for forming the EDE are large orbital
eccentricity, low fraction of the stellar wind mass
present within the orbital plane,
large stellar mass, small orbital separation,
and low mass-loss rate. Such conditions are often
encountered in oxygen-rich stellar winds. Conversely, carbon-rich winds often exhibit a spiral structure rather than EDE
because of stronger acceleration by carbon-bearing dust
\citep{2020Sci...369.1497D}. Therefore, the chance of detecting
EDE in carbon-rich envelopes should not be large, although there exist such detections \citep[e.g.,][]{2004ApJ...617.1142S}.
However, our detection, together with those of 
\citet{2000ApJ...528..861U} and
\citet{2009ApJ...692..402N,2012ApJ...759...61N},
suggest that the probability of 21\,$\mu$m sources
harbouring an EDE is high. 
 \citet{2000AstL...26...88K} posit that the overabundance of silicon in IRAS~23304+6147 may be caused by the hot bottom burning (HBB) process occurring in the hot layers of the convective envelope of a massive AGB star. This is a characteristic of J-type carbon stars \citep{2007AJ....134..214C}, suggesting that the central star
has an initial mass greater than 4\,M$_{\odot}$. 
According to the formula of \citet{2020Sci...369.1497D},
the likelihood of developing an EDE structure would increase fivefold with a fourfold increase in stellar mass.
Therefore, it is reasonable
to infer that the formation of the 21\,$\mu$m carrier
might be associated to the short-lived EDE in massive carbon-rich PPNe. This could provide a
plausible explanation for  the low detection rate of this feature. 

On the other hand, HBB predominantly consumes $^{18}$O. 
\citet{2017A&A...600A..71D} show that the $^{17}$O/$^{18}$O ratio is 
positively correlated with the initial mass of the post-AGB star.
The $^{17}$O/$^{18}$O ratio approximately equals to
the ratio of  integrated intensities of  the C$^{17}$O and C$^{18}$O
lines.
We have detected the C$^{18}$O line; however, the C$^{17}$O line 
lies beneath our detection threshold.
Consequently, the upper limit of the $^{17}$O/$^{18}$O ratio is determined to be 0.5. In light of the relation presented by \citet{2017A&A...600A..71D},
it can be inferred that the stellar initial mass is less than 1.5\,M$_\odot$.
This inference implies that HBB does not play a  significant role in
this PPN, which stands in contrast to the conclusion drawn by
\citet{2000AstL...26...88K}. 
We put forward the hypothesis that the discrepancy might be attributable to binary evolution, even though the specific details remain obscure.


The EDE has been also detected in the mid-infrared image
of  another 21\,$\mu$m PPN IRAS~07134+1005 \citep{2000ApJ...528..861U,2002ApJ...573..720K}, which traces the dust emission. However, high-angular resolution
CO and $^{13}$CO maps show that it is a hollow shell with a slight velocity enhancement along the major axis, casting
a doubt on the presence of the EDE \citep{2019IAUS..343..520U}. 
If looking at the CO map along (Figure~\ref{co_channel}),
we may have made the same claim  for the morpho-kinematics of IRAS~23304+6147 that
no EDE is seen. However, as clearly shown in Figure~\ref{compare_molecule}, carbon-chain molecules trace a torus-like structure. The CO map displays significantly different 
morphology  mostly due to its optically thick nature. On the other hand, C$_4$H and HC$_3$N  have subtly different
spatial distributions with the former more concentrated in the
torus, probably suggesting a chemical differentiation. 
From above discussion we can learn that PPNe's
appearances have a molecule dependence. Utilizing the images of only one or 
two molecular lines may result in an unrealistic  
classification of nebular morphologies. 
 \citet{2024Galax..12...63M} provide compelling evidence for this molecular selectivity. Their 
 interferometric observations demonstrate that 
HCO\(^+\) and HCN maps  resolve knotty structures 
 spanning the central region to the polar tips of
 M1-92. Notably, these structures are absent in both the CO and $^{13}$CO  maps.
Future observations with high spatial resolution could reveal the distribution of a wider range of molecules, not just CO, within  the EDE and EE post-AGB structures.

\subsection{Chemistry of the 21\,$\mu$m PPN} \label{disb}

The departure from  spherical symmetry could result in
spatially distinct chemical differentiation.
The parent molecules ejected from the AGB star are ionized
or dissociated by the UV photons from the external 
interstellar radiation field, which induces ion- or 
radical-neutral reactions, and thus increases the
molecular complexity in the circumstellar envelope.
As the star evolve into the PPN phase, the increasing temperature of the central star and the expansion of the  envelope may strengthen the 
internal UV radiation, 
which in principle could also contributes to the photo and
radiation chemistry of the envelope. 
State-of-the-art astrochemical models predict that 
under the assumption of a clumpy stellar wind and additional
radiation from a hot companion, sufficient internal UV photons can 
penetrate deeply
into the envelope and impose a significant influence on the circumstellar
chemistry \citep{2020ChJCP..33..668M,2023FaDi..245..586V}.
Observational evidence for circumstellar chemistry
driven by the internal UV radiation  is 
still scarce.
Different chemical processes in the EE and EDE  may occur due to their different densities and extinctions that govern
the penetration of the UV photons into the envelope.
In the low-density EE, the collisions between molecules are relatively inefficient and the UV radiation is intense, and thus the synthesis of daughter molecules from photodissociation products is inefficient. The situation is contrary for the EDE, where the density is higher and daughter molecules can be synthesized more efficiently.
With this in mind, the following subsections attempt to provide
an interpretation of the spatial distribution of molecules in this PPN.  In Sections~\ref{dis421} and \ref{dis422}, we discuss the carbon- and silicon-bearing molecules, respectively.
Section~\ref{dis423} presents the spatial distribution of
the $^{12}$CO/$^{13}$CO ratio.

\subsubsection{Carbon-bearing molecules} \label{dis421}

\begin{figure*}
    \centering
    \subfigure{
        \includegraphics[width=0.428\linewidth]{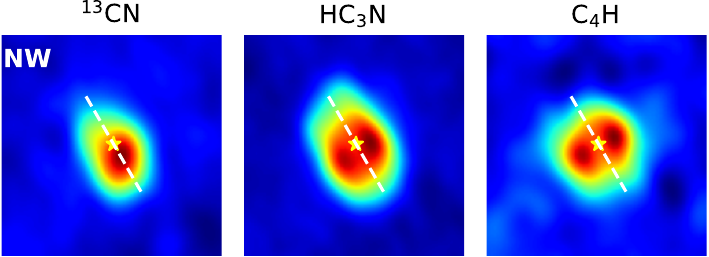}
        \label{fig:outflow_dis_c}
    }
    \subfigure{
        \includegraphics[width=0.428\linewidth]{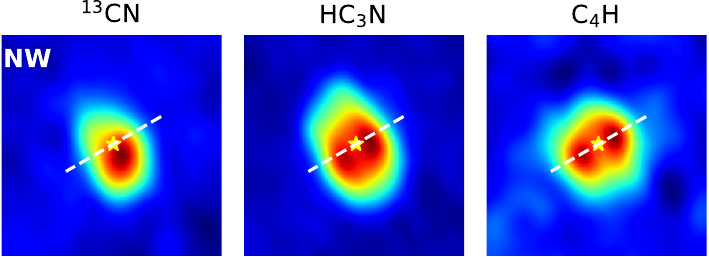}
        \label{fig:outflow_slit}
    }
    \subfigure{
        \includegraphics[width=0.48\linewidth]{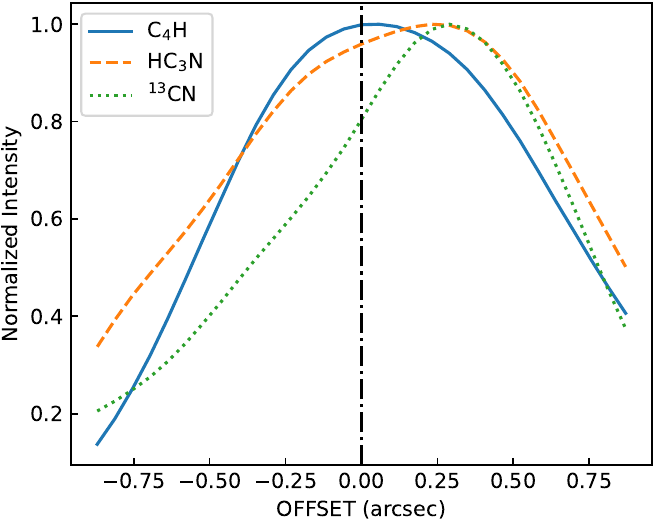}
        \label{fig:torus_slit}
    }
    \subfigure{
        \includegraphics[width=0.48\linewidth]{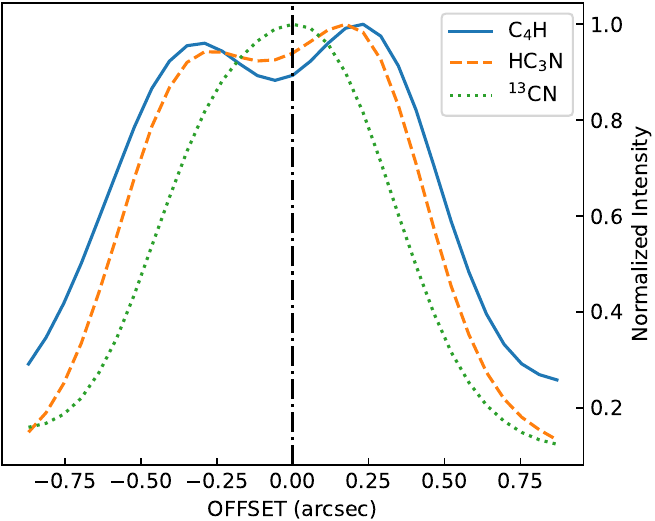}
        \label{fig:trous_dis_c}
    }
  \caption{Spatial distributions of $^{13}$CN, HC$_3$N, and C$_4$H molecules. The northwest direction is marked in the upper left corner of the $^{13}$CN map. The left panels show the molecular line intensity distributions along the major-axis direction,  
  tracking a path from northwest to southeast (with the offset changing from negative to positive) at a PA of  
  $30^\circ$ relative to the north-south direction. 
  The right panels 
 show the molecular line intensity distributions along the 
 minor-axis direction,   
  tracking a path from southwest to northeast (with the offset changing from negative to positive) at a PA of $60^\circ$. 
  In the upper panels, the dashed lines
  denote the paths along which the line intensity profiles are extracted. The positions of the phase centers are denoted by 
  the pentagrams in the upper panels and 
  the vertical dashed lines in the lower panels.
}
  \label{distribution_c}
\end{figure*}

\begin{figure*}
    \centering
\includegraphics[width=1\linewidth]{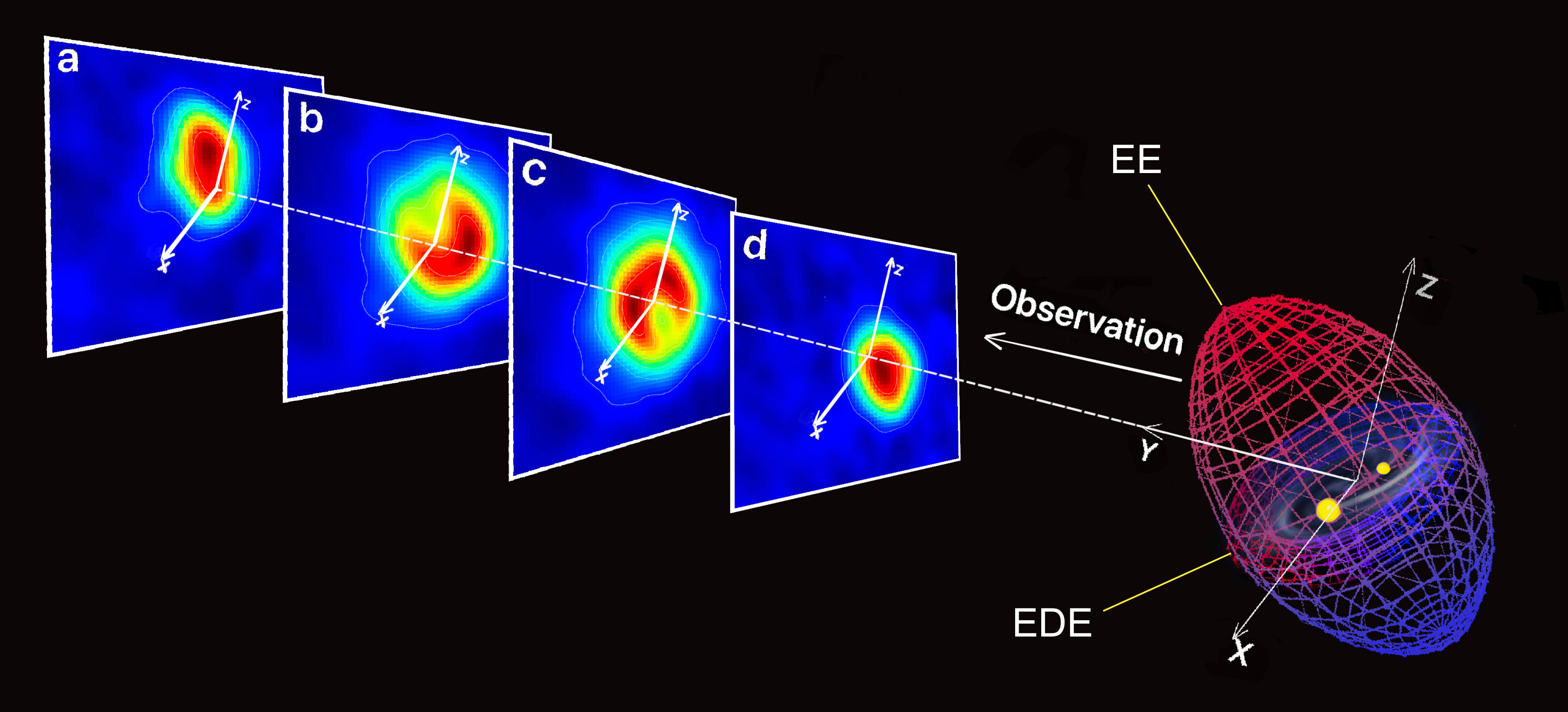}
    \caption{ 
    Schematic illustration of the true shape
    of IRAS~23304+6147.
    The grid on the right represents  
    the conceptual model consisting of a hypothetical binary star system (yellow filled circles), an EE, and an EDE. The coloration
    of the grid is defined according to the Doppler effect, that is,  the regions moving away from the observer's perspective (along the Y-axis, as denoted by the arrow below `Observation') are indicated in red, while those moving towards the observer in blue.
    Four representational velocity slices
    are shown on the left: a ($V_{\rm LSR}=-3$ to $3$\,$\rm km~s^{-1}$), b ($V_{\rm LSR}=-16$ to $-5$\,$\rm km~s^{-1}$), c ($V_{\rm LSR}=-27$ to $-18$\,$\rm km~s^{-1}$), and d ($V_{\rm LSR}=-40$ to $-29$\,$\rm km~s^{-1}$). The nebula is
    projected onto the X-Z plane (the sky plane). 
    From the observer's perspective, from the far to near sides, the EE is first seen in channel `a' and progresses towards  `d', while the EDE shifts from channel `b' to `c'.}
    \label{model}
\end{figure*}

For the sake of clarity, Figure~\ref{distribution_c} presents
the line intensity profiles of three carbon-bearing molecules
along the major axis of the EE and the EDE (i.e., the minor axis of the EE). As shown in the right panels of this figure, larger  carbon-bearing molecules are located farther out in the EDE.
HC$_3$N and C$_4$H exhibits a double-peak profile with the latter
lying in the extended region of the EDE, while $^{13}$CN  shows a single peak. 
Along the direction of the EE major-axis (the left panels
of Figure~\ref{distribution_c}), the intensity profiles of $^{13}$CN  and HC$_3$N lines
are asymmetric with the peak intensities slightly shifting to  the southeast. This  can be attributed to the obscuration of
the emission from the EE by the inclined front torus (see Figure~\ref{model}).
The asymmetry is more pronounced for $^{13}$CN, suggesting that 
this molecule is located in the innermost region of the EE.
C$_4$H lines exhibit a symmetrical intensity profiles, indicating
that they are located predominantly in the outer EDE.
The HC$_3$N line profile appears to fall somewhere in between those of the $^{13}$CN and C$_4$H, and could be decomposed into two components respectively corresponding to the EE and EDE.

The breadths ($R$) of the spatial distributions of the molecular emission along the EDE diration have a relation of $R(^{13}{\rm CN})<R({\rm HC_3N})<R({\rm C_4H})$. The dipole moments ($\mu$) of the three carbon-bearing molecules follow $\mu(^{13}{\rm CN})<\mu({\rm C_4H})<\mu({\rm HC_3N})$. The HC$_3$N  and C$_4$H lines have a similar excitation energy
that is substantially higher than that of the $^{13}$CN lines
(Table~\ref{line}). Therefore, it is hard to explain the
different spatial distributions of the molecular emission 
in the EDE in terms of different excitation conditions,
which are more likely to stem from chemistry.

In the gas-phase chemistry, HC$_3$N is primarily formed through 
the reaction between C$_2$H$_2$ and CN
\citep{2017A&A...601A...4A}. The gas ejected from
the stellar surface is rich in C$_2$H$_2$ and HCN. CN is available
only when HCN is dissociated by sufficiently intense UV radiation.
According to  traditional astrochemical models, UV photons are
mainly from the interstellar medium, and thus CN and HC$_3$N
should  be enriched in the outer regions of the envelope with CN lying more outside \citep[see e.g.,][]{1994A&A...288..561M,2017A&A...601A...4A}. This is opposite to
our observations, providing a strong evidence for the photochemistry driven by the internal UV radiation.
The photochemistry of the cyanopolyyne family  has also been 
revealed in CRL\,618 \citep{2005ApJ...628..275P}. However, CRL\,618 has
a central B0 star that has developed a compact ionized region,
and thus is now leaving the PPN stage. Unlike that of
CRL\,618, the central stars of 21\,$\mu$m sources have a temperature of $\sim5000$\,K; their photochemistry may be driven by an obscured hot companion.

There are two potential routes to form circumstellar C$_4$H. First, this species can be formed via C$_2$H$_2+$C$_2$H, where C$_2$H can be enriched
 by the photodissociation of HC$_3$N
 \citep{2017A&A...601A...4A}. If this is the case,
 the distribution of C$_4$H provides a further supporting
 arguement for the hypothesis of an internal photochemistry. 
 Alternatively, C$_4$H can be formed via C$_2$H$_2+$C$_2$ which is efficient
 even at low temperatures \citep{2007Icar..187..558C}.
The outer midplane of the EDE is short of photon heating, and thus is colder,
providing a favourable environment
for the formation of C$_4$H.
This is consistent with the observed spatial distribution of C$_4$H.
The rate coefficients of these neutral-neutral reactions are typically
$10^{-10}$\,cm$^3$\,s$^{-1}$. If assuming the density of 
C$_2$H$_2$ to be 1\,cm$^{-3}$, the reaction timescale is
$\sim300$\,years, which is comparable to the dynamical timescale
of the EDE. Since large hydrocarbons can be formed in the EDE,
it is reasonable to speculate that it also harbours more complex organics. The enhancement of hydrocarbon
clusters in 21\,$\mu$m sources have been taken as an indication of carbonaceous molecules as the carrier of the
21\,$\mu$m feature
\citep{1990ApJ...365L..23B}.






\subsubsection{Silicon-bearing molecules}  \label{dis422}

\begin{figure*}
    \centering
\includegraphics[width=1\linewidth]{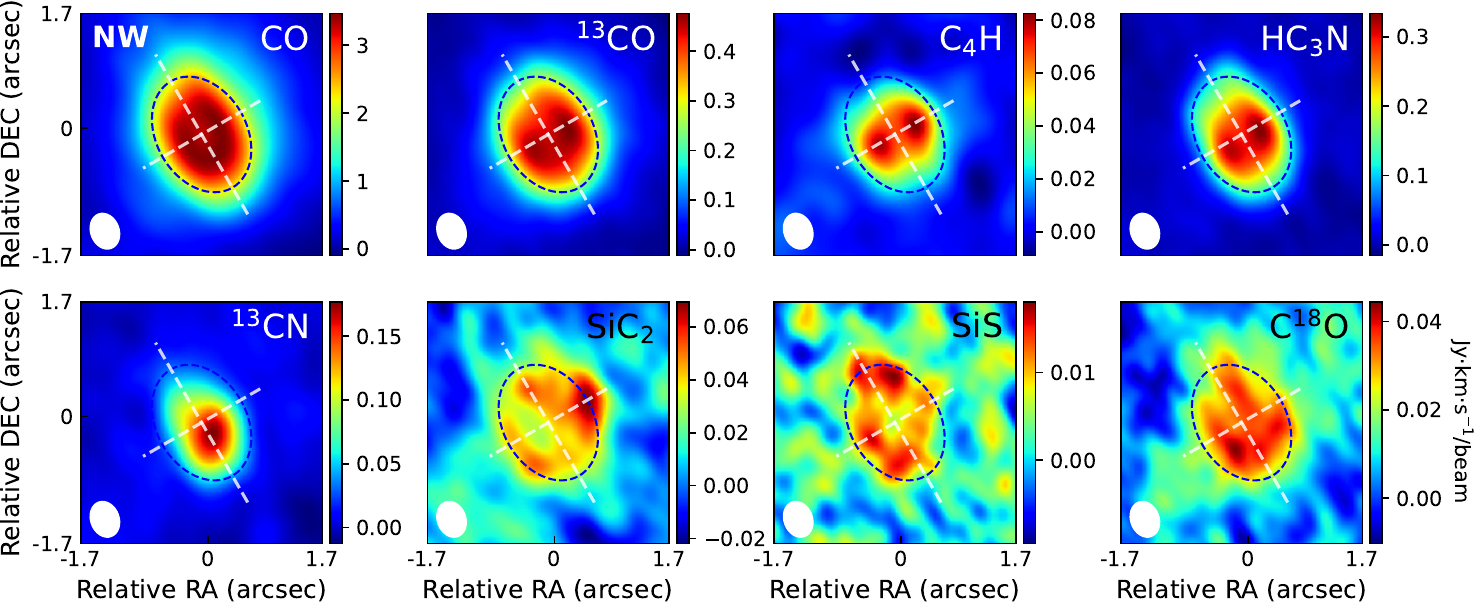}
    \caption{Integrated intensity (moment 0) maps of the molecular lines detected in IRAS~23304+6147.
    The white ellipse at the down-left corner represents the synthesized beam.
    For comparison convenience, in each panel the dashed elliptical
    circles indicate the main CO emission region that have
    a semi-major (PA$=30^\circ$) and minor axis 
    (PA$=60^\circ$) of 1.16$\arcsec$ and 0.84$\arcsec$,
    and the dashed straight lines mark the size of
    the extended CO halo. The northwest (NW) direction 
    is marked in the upper-left corner of the CO  map. }
\label{compare_molecule}
\end{figure*}

\begin{figure*}
    \centering
\includegraphics[width=1\linewidth]{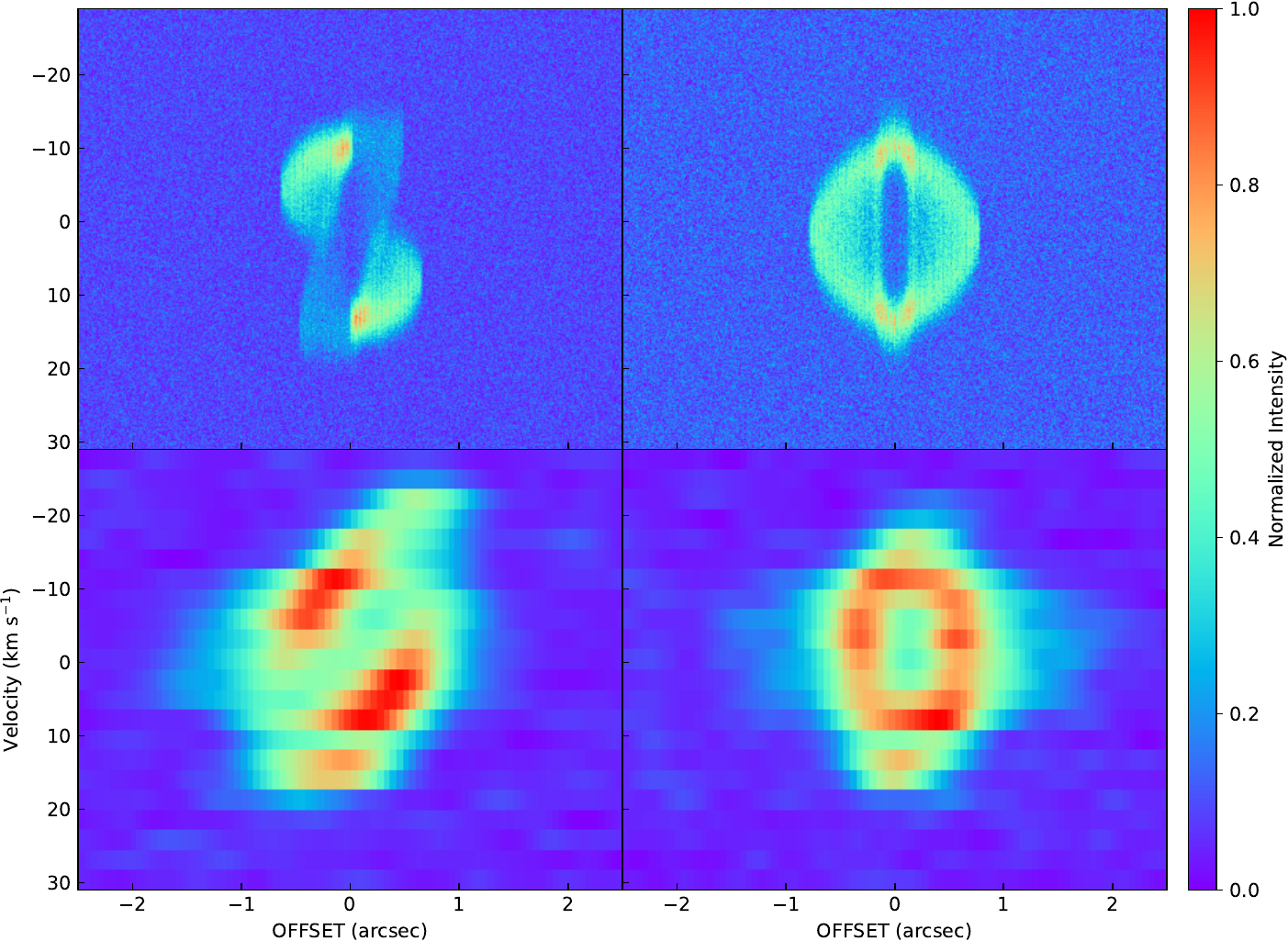}
    \caption{PV diagrams of the
    $^{13}$CO $J=2$--1 transition, derived from the modelling
    (upper panels) and observations (lower panels).      
    The left panels present the velocity structure along the major axis of the EE, oriented from northwest to southeast, where negative offsets indicate the northwest direction. The right panels show the velocity structure along the minor axis, oriented from southwest to northeast, with negative offsets corresponding to the southwest direction. The cuts used to extract these diagrams are detailed in Figure~\ref{compare_molecule}. }  
    \label{pv_model}
\end{figure*}

Differing from carbon-bearing molecules, silicon-bearing molecules are
distributed around the periphery of the EDE
(Figure~\ref{compare_molecule}). Given their refractory nature,
silicon-bearing molecules  can readily condense onto dust particles.
\citet{2018A&A...611A..29M} find that the SiC$_2$ abundance
in carbon-rich AGB envelopes strongly decrease with increasing  density, and infer that SiC$_2$ is the gas-phase precursor of
SiC dust. The absence of gaseous 
silicon-bearing molecules in the EDE may suggest 
an efficient dust formation.
This conforms with the mid-infrared detection of the EDE in
IRAS~07134+1005 \citep{2000ApJ...528..861U,2002ApJ...573..720K}.
The synthesis of circumstellar SiC$_2$ has been a long-standing debate, centered on whether it is a parent molecule existing in the photosphere or a daughter species formed in the outer envelope through ion recombination reactions of parent molecules \citep{1990MNRAS.244..444H,2023FrASS..1015642F}. 
As shown in Figure~\ref{compare_molecule}, SiC$_2$ and SiS seem
to exhibit complementary spatial distributions, suggesting that
the two species are chemically linked.  SiC$_2$ can be formed from SiS through the reaction routes $\mathrm{SiS}\rightarrow\mathrm{SiC_2H^+}\rightarrow\mathrm{SiC_2}$.

Notably, a broad emission around 30\,$\mu$m  is consistently detected in sources exhibiting the 21\,$\mu$m feature, suggesting a profound connection between the two features \citep{2020Ap&SS.365...88V}.
The 30\,$\mu$m feature is widely attributed to
MgS \citep{1985ApJ...290L..35G, 1985ApJ...290L..41N}.
\citet{2008A&A...486..229Z} present that
in carbon-rich environments 
MgS can be formed by reactions between
Mg and 
gas-phase sulfur-bearing molecules only when
Si is  heavily
depleted into solid SiC. Otherwise S is primarily locked in SiS.
Therefore, the severe condensation of Si-bearing molecules in the 
EDE may have boosted the formation of MgS. As a results, the
30\,$\mu$m feature is strong in this source.
When the dusty EDE is exposed to the internal UV radiation,
there may be a wealth of a dust-gas chemistry \citep{2023FaDi..245..586V}.
Complex organic molecules, such as hydrogenated fullerenes
\citep{2020ApJ...898..151Z},  may be efficiently formed and be responsible for the 21\,$\mu$m  feature.
This hypothesis is consistent with the findings of \citet{2002ApJ...573..720K} that the aromatic/aliphaic and 21\,$\mu$m features and the dust emission arise from the same location.


\subsubsection{$^{12}$CO/$^{13}$CO ratio}  \label{dis423}

It is well established that the isotopic $^{12}$C/$^{13}$C  ratios in evolved stars are lower than the solar value (89).
According to \citet{2014A&A...566A.145R}, the  median values of the $^{12}$CO/$^{13}$CO abundance ratios in M-, S-, and C-type AGB stars are 13, 26, and 34, respectively.
The CN cycle in the hydrogen-burning shell
converts $^{12}$C into the $\beta$-unstable isotope
$^{13}$N, which decays into $^{13}$C. 
The third dredge-up transports $^{13}$C to the stellar surface, lowering circumstellar $^{12}$C/$^{13}$C ratio. 
In binary stellar systems, the $^{13}$C enhancement could be
caused by nova pollution \citep{Sengupta2013ANR}.
The $^{12}$CO and $^{13}$CO ($J=2$--1) lines in IRAS~23304+6147 have an integrated intensity ratio of 7.1. 
The value is similar to the mean value of other  21\,$\mu$m sources \citep[$10.3\pm3.1$,][]{2005ApJ...624..331H,2024AJ....167...91Q},
suggesting that 21\,$\mu$m sources have a low 
$^{12}$C/$^{13}$C abundance ratio.
Such low $^{12}$CO/$^{13}$CO ratios have also been 
found in binary post-AGB stars with (sub-)Keplerian rotating disks
\citep[$\sim8.6$,][]{2022A&A...659A.134G}.
J-type carbon stars have  relatively low  $^{12}$C/$^{13}$C ratios
which might be related to their binary nature 
\citep[e.g.,][]{izzard07,Sengupta2013ANR,2000ApJ...536..438A}, although the exact cause is unclear.
It is reasonable to hypothesize that the 21\,$\mu$m  sources might be descended from
J-type carbon stars.

\begin{figure*}
    \centering
\includegraphics[width=0.7\linewidth]{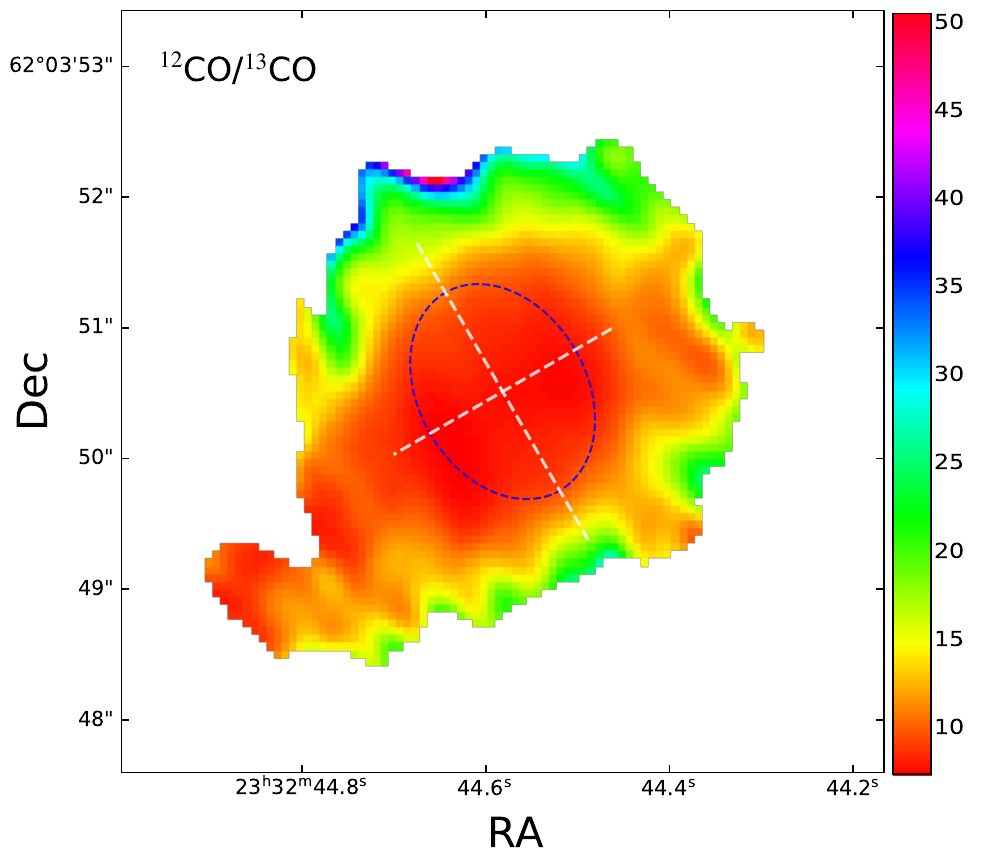}
    \caption{Ratio map between $^{12}$CO and $^{13}$CO
    integrated intensities. 
     The dashed ellipse is
    the same as that in Figure~\ref{compare_molecule}. The major
    and minor axes of the ellipse align with the EE and EDE direction, respectively.}  
    \label{12c_13c}
\end{figure*}

Figure~\ref{12c_13c} presents the $^{12}$CO/$^{13}$CO map,
demonstrating the notable differences between the EDE and EE.
The ratio is comparatively low (6.9--15) in the EDE and larger 
(20--50) in the EE, peaking in the outermost regions. 
This might be because the EDE's optical thickness is higher than the EE's. On the other hand, the outermost regions of the EE
are exposed to the interstellar radiation field, causing
 photodissociation of $^{13}$CO. As $^{12}$CO may remain largely intact due to its self-shielding effect, the $^{12}$CO/$^{13}$CO
 is larger in the tips of the EE. As shown in Figure~\ref{compare_molecule}, the 
$^{12}$CO emission has a larger extent along the EE major axis
than the $^{13}$CO emission. Therefore, $^{13}$CO is unable
to fully represent the actual nebular structure. 
We propose that the EE and EDE may constitute structural components of an oval-shaped nebula.
In Figure~\ref{model}, we present a schematic illustration of the
intrinsic geometry of this PPN, with the diffuse halo omitted
for visual clarity.

\section{Conclusion} \label{sec:concl}

We present interferometric NOEMA maps of gas-phase molecules in the 21\,$\mu$m source IRAS~23304+6147,  which offer valuable information about the structure of this nebula. Eight 
different molecules and isotopologues are observed, among 
which six (C$^{18}$O, SiS, SiC$_4$, $^{13}$CN, HC$_3$N, and C$_4$N) are new detections in this object. 
In comparison to the 
extensive research on the chemistry of 
AGB  envelopes,  the chemistry of PPNe has
been barely explored. The newly detected molecular lines have, to some extent, deepened our understanding of the molecular chemistry in these objects. We constructed a morpho-kinematic model to reproduce the
observed maps.
A chemically stratified structure is revealed, showcasing two distinct features: EDE and EE. Based on the $^{13}$CO emission, we 
estimate that the nebular mass is larger than $1.3\times10^{-2}$\,M$_\odot$. A comparison with single-dish observations indicates that an extended halo might be missed
by our interferometric observations.
Linear carbon-chain molecules  preferentially  distribute in the EDE, forming a torus-like structure. The different spatial distributions of $^{13}$CN,
HC$_3$N, and C$_4$H provide an strong evidence that the internal
UV radiation initiates the photochemistry. SiS and SiC$_2$ maps
exhibit a hollow, indicating to an efficient dust formation in the EDE. These results show that the transition of the PPN 
morphology may result in significant chemical differentiation.
As a result, different species may has distinct spatial 
distributions, and the morpho-kinematics of PPNe have a molecule
dependence.

The $^{12}$CO/$^{13}$CO ratios span from 6.9 in the core of the EDE 
to  50 in the tips of EE and stay approximately constant in the main nebular regions. This suggests  that the EDE is optically thick in $^{12}$CO and that $^{13}$CO has been significantly photodissociated in the outermost regions. The generally low
 $^{12}$CO/$^{13}$CO ratio and the presence of EDE structure
 in IRAS~23304+6147 seem to indicate that its progenitor is
 probably a J-type carbon star within a binary system.

Such a carbon-rich EDE may hold the vital clue to
the origin of the 21\,$\mu$m feature. Our observations show that 
the EDE serves as a favourable environment for dust formation and molecular synthesis. Speculatively, complex aromatic/aliphatic 
compounds can be formed in the EDE, some of which, such
as hydrogenated fullerenes, might be
the carrier of the 21\,$\mu$m feature.

\begin{acknowledgements}

We thank the anonymous reviewer for insightful suggestions, which have significantly improved the quality of this work.
We are grateful to 
Ana Lopez Sepulcre and Michael Bremer for their helps in the data reduction. 
The financial supports of this work are from 
the National Natural Science Foundation of China (NSFC, No. 12473027,  12333005, and 11973099),
the Guangdong Basic and Applied Basic Research Funding (No.\,2024A1515010798),
and the science research grants from the China Manned Space Project (NO. CMS-CSST-2021-A09, CMS-CSST-2021-A10, etc).
Y.Z. thanks the Xinjiang Tianchi Talent Program (2023). 
X.H.L. acknowledges support from the Natural Science Foundation of Xinjiang Uygur Autonomous Region (No. 2024D01E37) and the National Science Foundation of China (12473025).
This work is based on observations carried out 
with the IRAM NOEMA Interferometer. IRAM is supported by INSU/CNRS (France), MPG (Germany) and IGN (Spain). 
\end{acknowledgements}
%




\bibliographystyle{aa}

\newpage

\begin{table*}
\centering
\caption{Molecular lines detected in the observations.}
\begin{tabular}{ccccccc}
\hline \hline
Molecule & Transition &\begin{tabular}{c} E$_u$  \\ (K) \end{tabular}& \begin{tabular}{c} Frequency \\ (MHz) \end{tabular}  & \begin{tabular}{c} $I_{\text{peak}}$ \\ (K) \end{tabular} & \begin{tabular}{c} $\int I \, dv$ \\ (K  $\rm km~s^{-1}$) \end{tabular} & \begin{tabular}{c} rms  \\ (K) \end{tabular} \\
\hline
CO & J=2-1 & 16.59 & 230538.00  & 22.17 & 216.80 & 0.16 \\
$^{13}$CO & J=2-1& 15.86 & 220398.67  & 3.14 & 30.65 & 0.14 \\
HC$_3$N & J=24-23 & 130.98& 218324.71 & 2.03 & 18.85 & 0.14 \\
SiC$_2$ & J$_{Ka,Kc}=9_{4,6}-8_{4,5}$ & 82.28& 213208.03  & 0.32 & 2.04 & 0.15 \\
 & J$_{Ka,Kc}=9_{4,5}-8_{4,4}$ & 82.28& 213292.33  & 0.32 & 2.01 & 0.15 \\
 & J$_{Ka,Kc}=10_{2,9}-9_{2,8}$ & 69.62& 232534.07  & 0.28 & 1.97 & 0.18\\
  & J$_{Ka,Kc}=10_{6,5}-9_{6,4}$ & 132.37& 235713.00  & 0.25 & 1.85 & 0.18 \\
 C$_4$H & N,J=23-22, 47/2-45/2 & 126.03& 218836.98  & 0.38 & 3.15 & 0.14 \\
 & N,J=23-22, 45/2-43/2 & 126.07& 218875.35  & 0.45 & 3.56 & 0.14 \\
 &N,J=24-23, 47/2-45/2 & 137.03& 228386.92  & 0.40 & 3.11 & 0.16 \\
 C$^{18}$O & J=2-1 & 15.81& 219560.35  & 0.32 & 2.81 & 0.14 \\
 SiS & J=12-11 & 67.95& 217817.66 & 0.18& 0.58 & 0.14 \\
$^{13}$CN 
& N,J,F=2-1,3/2-1/2,2-2 & 15.66& 217074.23 &0.15& ... & 0.14 \\
& N,J,F=2-1,5/2-3/2,3-2 & 15.65& 217303.19 &0.71 & ... & 0.14 \\
& N,J,F=2-1,3/2-1/2,2-1 & 15.66& 217304.92 &0.72 & 5.12 & 0.14 \\
& N,J,F=2-1,5/2-3/2,2-2 & 15.68& 217437.70 & 0.38& 2.10 & 0.14 \\
& N,J,F=2-1,5/2-3/2,4-3 & 15.68& 217467.15  & 0.52& ... & 0.14 \\
& N,J,F=2-1,5/2-3/2,2-1 & 15.69& 217469.15 & 0.52& 4.10 & 0.14 \\
& N,J,F=2-1,5/2-1/2,2-1 & 15.68& 217668.43 & 0.17 & 1.13 & 0.14 \\
& N,J,F=2-1,5/2-1/2,1-1 & 15.68& 217675.39 & 0.17& 1.11 & 0.14 \\
\hline
\end{tabular}
\tablefoot{ The coefficient to convert the intensity unit form Jy\,beam$^{-1}$ to K is 
105--122\,K/(Jy\,beam$^{-1}$) at 100--122\,GHz for
a  synthesized beam size of $0.52\arcsec\times0.42\arcsec$.}
\label{line}
\end{table*}

\clearpage

\end{document}